\documentclass[]{jfm}

\usepackage{graphicx}
\usepackage{newtxtext}
\usepackage{newtxmath}
\usepackage{natbib}
\usepackage{graphicx}
\usepackage{tabulary}
\usepackage{subcaption}
\usepackage{fontawesome}
\usepackage{epstopdf, epsfig}
\usepackage[dvipsnames]{xcolor}
\usepackage{color}
\definecolor{matlab-green}{rgb}{0,0.5,0}
\definecolor{matlab-purple}{rgb}{0.4940, 0.1840, 0.5560}

\newcommand{\black}{\color{Black}}
\usepackage{hyperref}
\hypersetup{
    colorlinks = true,
    urlcolor   = blue,
    citecolor  = blue,
}

\newcommand{\RomanNumeralCaps}[1]
\linenumbers

\usepackage{breqn}

\title{A vortex sheet based analytical model of the curled wake behind yawed wind turbines}

\author{Majid Bastankhah\aff{1} \corresp{\email{majid.bastankhah@durham.ac.uk}},  Carl R. Shapiro\aff{2,3},
 Sina Shamsoddin\aff{4},
Dennice F. Gayme\aff{2}
 \and Charles Meneveau\aff{2}
\affiliation{\aff{1} Department of Engineering, Durham University, Durham DH1 3LE, United Kingdom
\aff{2} Department of Mechanical Engineering, Johns Hopkins University, Baltimore, MD 21218, USA
\aff{3} AAAS Science and Technology Policy Fellow, Building Technologies Office, U.S. Department of Energy, Washington, DC 20585, USA
\aff{4} Swiss Finance and Property Group, Seefeldstrasse 275, 8008 Zurich, Switzerland} }

\begin{document}

\maketitle

\begin{abstract}
Motivated by the need for compact descriptions of the evolution of non-classical wakes behind yawed wind turbines, we develop an analytical model to predict the shape of curled wakes. Interest in such modelling arises due to the potential of wake steering as a strategy for mitigating power reduction and unsteady loading of downstream turbines in wind farms. We first estimate the distribution of the shed vorticity at the wake edge due to both yaw offset and rotating blades. By considering the wake edge as an ideally thin vortex sheet, we describe its evolution in time moving with the flow. Vortex sheet equations are solved using a power series expansion method, and an approximate solution for the wake shape is obtained.
The vortex sheet time evolution is then mapped into a spatial evolution by using a convection velocity. 
Apart from the wake shape, the lateral deflection of the wake including ground effects is modelled. Our results show that there exists a universal solution for the shape of curled wakes if suitable dimensionless variables are employed.  For the case of turbulent boundary layer inflow, the decay of vortex sheet circulation due to turbulent diffusion is included. Finally, we modify the Gaussian wake model by incorporating the predicted shape and deflection of the curled wake, so that we can calculate the wake profiles behind yawed turbines. Model predictions are validated against large-eddy simulations and laboratory experiments for turbines with various operating conditions. 
\end{abstract}


\section{Introduction}
Analytical models of various fluid mechanical phenomena in wind energy play an important role for basic understanding and for design and control of wind farms. Prime examples are models for the wind turbine wake velocity defect and its downstream evolution commonly used in wind farm layout optimization \citep{Jensen1983,stevens2017review,porte2020}. In the classic Jensen model, for instance, a linearly expanding wake  with a top-hat shape is assumed. More recent models allow for more realistic wake cross-sections with a Gaussian distribution \citep{bastankhah2014new}, and cross sections that transition from top-hat near the turbine to Gaussian further downstream  have also been proposed \citep{shapiro2019paradigm}. 

Analytical wake models can be particularly useful in implementation of wake mitigation strategies such as wake steering, which has been receiving growing attention as an important control approach for increasing wind farm power output \citep{fleming2014evaluating,gebraad2014wind,bastankhah2015wind,Howland2016,campagnolo2016,schottler2017,Bartl2018,Lin-porteagel2019,kleusberg2020,speakman2021}. Achieving increased power output through wake steering  involves turbines, often in the front rows of wind farms, being intentionally operated in yawed conditions to redirect their wakes away from downwind turbines. Although this reduces the power produced by the yawed turbines, research has shown that the total wind farm efficiency can improve as a result of more power generated by downwind turbines \citep{Park2016a,bastankhah2019wind,fleming2019,howland2019wind,King2021secondary}. 
Yawed turbine wake flows are known to exhibit complex features  which makes their modelling more challenging than their  unyawed counterparts. The most striking fluid-dynamic feature of a yawed turbine wake is arguably its \emph{curled} cross-sectional shape (i.e., a kidney-shaped cross-section). This shape arises  due to the action of a counter-rotating vortex pair (CVP) as detailed in \cite{Howland2016}. CVPs are typically generated when forcing with strong spanwise gradients is applied perpendicular to the flow direction. One of the most notable examples are vortices trailing from finite-span wings that roll up and lead to the formation of a CVP, i.e. wingtip vortices or wake vortices in the aerodynamics literature. 
The formation and evolution of these vortical structures have been the subject of numerous studies since seminal works of Prandtl and Lancaster \citep[See][and references therein]{Anderson2011Fundamentals}.
As noted in  \citet{bastankhah2016experimental}, the CVP observed in yawed turbine wakes is also similar to those formed in many other free shear flows with strong spanwise variations of cross-wind velocity such as cross-flow jets (see e.g., the review of \citet{mahesh2013}). 
 

In order to exploit the basic understanding of induced velocity and circulation of CVPs generated by  finite-span wings, \citet{shapiro2018} proposed to regard a yawed turbine as a lifting surface with an elliptical planform. Based on this approach, the lateral component of the thrust force can be regarded as the transverse lift force. This analogy made it possible to determine: (i) distribution of circulation at the yawed rotor modelled as a lifting line, and (ii) transverse velocity (equivalent to \emph{downwash} velocity for finite-span wings) at the rotor disk due to the yaw offset. The latter enabled modeling of the transverse displacement of the wake but the wake itself was still assumed to retain a circular cross-sectional shape rather than the curled shape observed in practice.  The associated vorticity distribution was later used by \citet{martinez2019aerodynamics,martinez2021} to develop a  Lagrangian vorticity transport model that can predict the curled shape of the wake after numerical integration. 
Recently,  \citet{Martinez2020equiva} and \citet{zong2020} have instead expressed rates of vorticity shedding at rotor blade tips using vortex cylinder theory \citep{coleman1945, Burton1995, branlard2016}  to determine the trailing vorticity distribution behind a yawed rotor. The numerical model developed by \citet{zong2020} also takes into account the redistribution of point vortices in the wake due to their self-induced velocities. More recently, \citet{shapiro2020decay} have solved the linearised mean streamwise vorticity transport equation to develop an analytical expression that can predict the decay of the CVP due to atmospheric turbulence.  \citet{bossuyt2021} have also experimentally demonstrated the impact of vortical structures shedding from a misaligned (either tilted or yawed) rotor on the curled shape of the wake downstream.


Capturing the curled shape of the wake for yawed turbines is of great importance since curling affects how much the wake will effectively overlap with downstream wind turbines, thus affecting the predicted power generation. However, models of the curled wake shape in the literature require numerical integration, and existing analytical wake models \citep[e.g.,][]{bastankhah2016experimental,shapiro2018,qian-ishihara2018,blondel2020yaw,King2021secondary} cannot capture this deformation of the wake shape. There are several advantages to analytically expressed models  that represent the trends in simple and explicit forms. 
 Apart from their low computational cost,   analytical flow models  \citep{meneveau2019}  often prove to be useful in revealing additional insights on flow physics that may not be evident using numerical simulation tools.  Therefore, the current study aims at developing an analytical model to predict displacement and shape deformation of the wake behind a yawed turbine. The proposed model is  inspired by prior works on two-dimensional vortex sheets \citep[e.g.,][]{rottman1987, coelho1989}. The proposed analytical model predicts displacement and deformation of a vortex sheet, shedding from the circumference of a yawed rotor, as it is convected downstream. The vortex sheet model is then combined with a  model for downstream evolution of wake velocity deficit to predict the shape of the curled wake and velocity distribution downwind of a yawed turbine. 

The remainder of the paper is organised as follows. Section \ref{sec:model_derivation} derives the vortex sheet, truncated power-series solution for the yawed turbine wake in uniform, ideal flow, and model predictions are compared with numerical simulation under laminar uniform inflow. In \S\ref{sec:turb_model}, the model is extended to cases with turbulent boundary-layer inflow, and the results are compared with corresponding  large eddy simulation (LES) data. Finally, \S\ref{sec:summary} provides a summary of the developed model and our main conclusions.

\section{Vortex sheet evolution in uniform, ideal flow}\label{sec:model_derivation}

\subsection{Evolving vortex sheet governing equations}\label{subsec:governing equation}
As shown in \citet{shapiro2020decay}, among others, vortices shedding from the yawed rotor circumference represented as an actuator disk form a tubular vortex sheet. 
The objective of this section is to model the shape evolution of this vortex sheet with downstream distance or, equivalently, with time. \black Only the streamwise component of the shedding vorticity is modelled in this work because the lateral wake deflection and the deformation of the wake cross-section are mainly due to the velocities induced by the streamwise component of shedding vorticity \citep{Martinez2020equiva}. The vortex sheet consists of semi-infinite streamwise vortex lines. In order to enable solving the governing equations  analytically, following \citet{coelho1989} and \citet{rottman1987}, we assume that the vortex sheet is planar and that its constituent vortex lines are infinite instead of semi-infinite, an approximation that improves at increasing distances from the origin. \black


\begin{figure}
\centerline{\includegraphics[width=\textwidth]{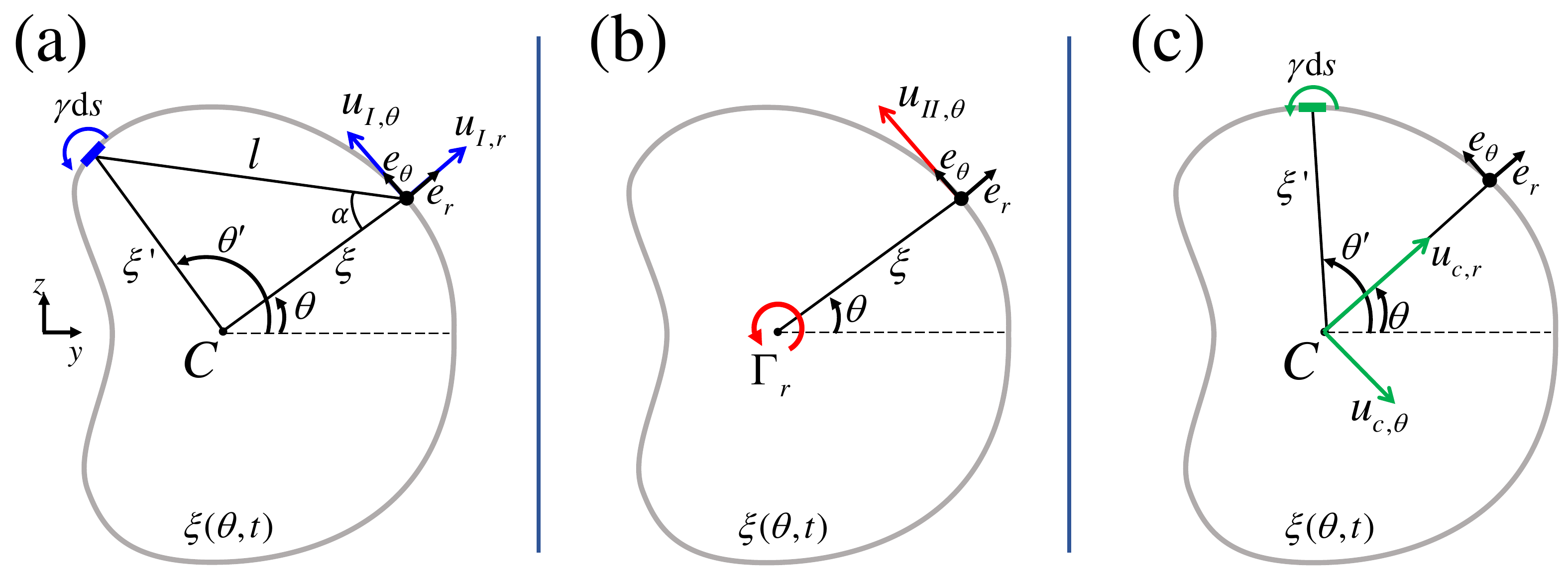}}
\caption{Schematic of the vortex sheet and different velocity terms on the right-hand side of (\ref{eq:vortex_sheet_velocity}). (a) Self induced vortex sheet velocity, $\mathbf{u}_I$. (b) Vortex sheet velocity induced by the point vortex at the vortex sheet centre, $\mathbf{u}_{I\!I}$. (c) velocity of the vortex sheet centre, $\mathbf{u}_{c}$. }\label{fig:schematic_vortex}
\end{figure}

Figure \ref{fig:schematic_vortex} shows a schematic of the vortex sheet in the plane normal to the incoming flow. The $(x,y,z)$ coordinate system is defined with an origin at the rotor centre and with $x$ in the streamwise direction (i.e., parallel to the incoming flow) and $y$ and $z$ in the spanwise and vertical directions, respectively. Alongside this Cartesian system, we define a polar coordinate system of $(r,\theta)$ in the $yz$-plane (i.e., plane normal to the incoming flow). This polar coordinate system is attached to the centre of the vortex sheet, denoted by $C$. The position of $C$ in the $(x,y,z)$ coordinate system is denoted by $(x_c,y_c,z_c)$. As $r$ is the radial distance from $C$, one can write $r^2=(y-y_c)^2+(z-z_c)^2$. The polar angle $\theta$ is measured from the positive $y$-axis toward the positive $z$-axis such that $\tan\theta=(z-z_c)/(y-y_c)$. The shape of the vortex sheet is represented by the polar function $\xi(\theta,t)$ that measures the distance of the vortex sheet from the centre, where $t$ is time. Our 
main objective is to describe the evolution of the vortex sheet as function of time $t$ in a frame moving downstream with the convection velocity $U_{con}$, i.e. determine $\xi(\theta,t)$. This is equivalent to determining the downstream spatial evolution of the vortex sheet, with $x=U_{con} t$.  
The sheet location $\xi(\theta,t)$ obeys 
\begin{equation}\label{Eq:xi_initial_eq}
    \xi=\xi_0 + \int u_r(\theta,t) \textrm{d}t
\end{equation}
where $\xi_0=\xi(\theta,0)$, and $u_r(\theta,t)$ denotes the radial velocity of the vortex sheet, which is affected by the strength of the evolving vortex sheet, whose evolution is treated next. 

Let us denote the strength of the vortex sheet by $\gamma=\gamma(\theta,t)$, where the vortex strength $\gamma$ is defined as the amount of circulation per unit length. In addition to the vortex sheet, there is a point vortex at $C$ with a circulation of $\Gamma_r$ (see figure \ref{fig:schematic_vortex}b) to model the rotor root vortex, which is elaborated later in \S\ref{subsect:initial_vorticity}. 
We will show in \S\ref{subsect:initial_vorticity} that the initial condition for the  strength can be written as
\begin{align}
       \label{Eq:gamma_0}
       \gamma_0&=\gamma(\theta,0)=\gamma_r+\gamma_b\sin\theta & \black \textrm{at  } & \black r=\xi_0, \black\\
          \label{Eq:Gamma_r}\Gamma_r&=-2\pi\xi\gamma_r & \black \textrm{at  } & \black  r=0\black,
\end{align}
where $\gamma_r$ and $\gamma_b$ are constants  depending on turbine operating conditions. Later in \S\ref{subsect:initial_vorticity}, we show that $\gamma_b$ is related to the vorticity generated due to  turbine yaw, while $\gamma_r$ and $\Gamma_r$ are related to the vorticity generated by turbine rotating blades. Our focus now is to predict the deformation of the vortex sheet provided that the initial conditions are given by (\ref{Eq:gamma_0}-\ref{Eq:Gamma_r}). 

The velocity of the vortex sheet $\textbf{u}(\theta,t)$ with respect to a coordinate system attached to the centre $C$ is given by

\begin{equation}\label{eq:vortex_sheet_velocity}
    \underbrace{\mathbf{u}(\theta,t)}_{\substack{\text{vortex sheet velocity} \\ \text{with respect to C}}}=\underbrace{\mathbf{u}_{\rm I}}_{\substack{\text{self-induced} \\ \text{vortex sheet velocity}}}+\underbrace{\mathbf{u}_{\rm II}}_{\substack{\text{vortex sheet velocity induced} \\ \text{by point vortex at C}}}-\underbrace{\mathbf{u}_{\rm c}}_{\substack{\text{velocity} \\ \text{of C}}}.
\end{equation}
In (\ref{eq:vortex_sheet_velocity}) and hereafter, bold letters denote vectors. Next, we employ the Biot-Savart law to determine the three velocity terms on the right-hand side of (\ref{eq:vortex_sheet_velocity}), starting with the self-induced velocity $\mathbf{u}_{\rm I}=u_{\textrm{ I},r}\mathbf{e}_r+u_{\textrm{I},\theta}\mathbf{e}_{\theta}$, where $\mathbf{e}_r$ and $\mathbf{e}_{\theta}$ are unit vectors in the radial and tangential directions, respectively. The radial $u_{\textrm{I},r}$ and tangential $u_{\textrm {I},\theta}$ components of the self-induced velocity at a given polar angle of $\theta$ are respectively given by
\begin{align}
       \label{Eq:u_r}
       u_{\rm I,r}(\theta,t)&=\int\limits_0^{2\pi}\frac{\gamma(\theta',t)\sin{\alpha}}{2\pi l}\xi'\textrm{d}\theta',\\
    \label{Eq:u_t}u_{\rmI,\theta}(\theta,t)&=\int\limits_0^{2\pi}\frac{\gamma(\theta',t)\cos{\alpha}}{2\pi l}\xi'\textrm{d}\theta',
\end{align}
where $l$ and $\alpha$ are defined in figure \ref{fig:schematic_vortex}a, $\theta'$ is a dummy integration variable, and $\xi'=\xi(\theta',t)$. According to the law of cosines, $
    l^2=\xi^2+\xi^{'2}-2\xi \xi'\cos{\left(\theta'-\theta\right)}.$
The angle $\alpha$ shown in the figure \ref{fig:schematic_vortex}a can be related to $l$ based on the law of sines for the drawn triangle, which results in $
    \sin{\alpha}=(\xi'/l)\sin{\left(\theta'-\theta\right)}.$
For small values of time $t$ and yaw angle $\beta$, we can assume that the vortex sheet is approximately circular and thus $\xi\approx\xi'$. After some trigonometric manipulations, (\ref{Eq:u_r}) and (\ref{Eq:u_t}) can be simplified to 
\begin{align} 
\label{Eq: u_I_r} 
u_{\rm I,r} (\theta,t) & = \mathrm{p.v.}\; \frac{1}{4\pi} \int_{0}^{2\pi} \frac{\gamma(\theta',t)}{\tan{\left[\left(\theta'-\theta\right)/2\right]}}\textrm{d}\theta', \\[3ex]
\label{Eq: u_I_t} 
u_{{\rm I}, \theta} (\theta,t) & = \mathrm{p.v.}\;\frac{1}{4\pi}\int_{0}^{2\pi} \frac{\gamma(\theta',t)\sin{\left[\left(\theta'-\theta\right)/2\right]}}{\sin{\left[\left(\theta'-\theta\right)/2\right]}}\textrm{d}\theta'=\frac{1}{4\pi} \int_{0}^{2\pi}\gamma(\theta')\textrm{d}\theta'.
\end{align}
Both integrals in (\ref{Eq: u_I_r}) and (\ref{Eq: u_I_t}) have singularities at $\theta'=\theta$ and thus we use the Cauchy principal values (p.v.) of these two integrals. While the p.v. of the latter can be simply obtained by removing $\sin[(\theta'-\theta)/2]$ from the numerator and the denominator, the p.v. of the former needs to be determined for a given $\gamma(\theta,t)$.

Next, we determine the velocity of the vortex sheet induced by the point vortex at C as shown in figure \ref{fig:schematic_vortex}b. 
We obtain
\begin{align} 
\label{Eq: u_II_r} 
u_{{\rm II},r} (\theta,t) &= 0, \\[3ex]
\label{Eq: u_II_t} 
u_{{\rm II},\theta} (\theta,t) & = \frac{\Gamma_r}{2\pi \xi}=-\gamma_r.
\end{align}
 Finally, we determine $\mathbf{u}_c$, which is the velocity of $C$ induced by the vortex sheet, shown in figure \ref{fig:schematic_vortex}c. It can be readily shown that  $\mathbf{u}_c$ is given by 
\begin{align} 
\label{Eq: u_c_r} 
u_{c,r}(\theta,t)&=\frac{1}{2\pi}\int\limits_0^{2\pi}\gamma(\theta',t)\sin\left({\theta'-\theta}\right)\textrm{d}\theta',\\
    \label{Eq:u_c_t}u_{c,\theta}(\theta,t)&=-\frac{1}{2\pi}\int\limits_0^{2\pi}\gamma(\theta',t)\cos\left({\theta'-\theta}\right)\textrm{d}\theta'.
\end{align}
If we neglect streamwise ($x$-direction) straining, vorticity is a conserved quantity, so the vorticity transport equation for the vortex sheet strength $\gamma(\theta,t)$ provides the additional required evolution equation \citep{moore1978}:
\begin{equation}\label{Eq:transport}
\frac{\partial \gamma}{\partial t}+ \frac{\partial\left(\gamma u_s\right)}{\partial s}=0,
\end{equation}
where $s$ is the arclength along the vortex sheet. At small values of time $t$ and yaw angle $\beta$, the vortex sheet remains approximately circular, so $u_s$ and $\partial s$ can be respectively replaced with $u_{\theta}$ and $\xi\partial\theta$. Therefore (\ref{Eq:transport}) is simplified to:

\begin{equation}\label{Eq:transp-final}
    \frac{\partial \gamma}{\partial t}+  \frac{1}{\xi}\frac{\partial (\gamma u_\theta)}{\partial \theta}\approx 0.
\end{equation}

Next, we use $\gamma_b$ and $\xi_0$ to non-dimensionalise variables in (\ref{eq:vortex_sheet_velocity}) and (\ref{Eq:transp-final}). This leads to a set of dimensionless equations as follows:
\begin{align} 
\label{Eq: u_r_dimensionless} 
\hat{u}_{r}(\theta,\hat{t})&\approx\mathrm{p.v.}\; \frac{1}{4\pi} \int_{0}^{2\pi} \frac{\hat{\gamma}(\theta',\hat{t})}{\tan{\left[\left(\theta'-\theta\right)/2\right]}}\textrm{d}\theta'-\frac{1}{2\pi}\int\limits_0^{2\pi}\hat{\gamma}(\theta',\hat{t})\sin\left({\theta'-\theta}\right)\textrm{d}\theta',\\
    \label{Eq:u_c_dimensionless}\hat{u}_{\theta}(\theta,\hat{t})&\approx\frac{1}{4\pi} \int_{0}^{2\pi}\hat{\gamma}(\theta')\textrm{d}\theta'+\frac{1}{2\pi}\int\limits_0^{2\pi}\hat{\gamma}(\theta',\hat{t})\cos\left({\theta'-\theta}\right)\textrm{d}\theta'-\upchi,\\
     \label{Eq:transport_dimensionless}    \frac{\partial \hat{\gamma}}{\partial \hat{t}} &\approx    -\frac{1}{\hat{\xi}} \frac{\partial (\hat{\gamma} \hat{u}_\theta)}{\partial \theta},
\end{align}
where $\hat{t}=t\gamma_b/\xi_0$, $\hat{u}=u/\gamma_b$,  $\hat{\gamma}=\gamma/\gamma_b$, $\hat{\xi}=\xi/\xi_0$ and $\upchi=\gamma_r/\gamma_b$. Note that the dimensionless time $\hat{t}$ becomes negative for negative values of $\gamma_b$.  In the following, we solve (\ref{Eq: u_r_dimensionless})-(\ref{Eq:transport_dimensionless}) using power series method.
\subsection{Analytical solution using power series approximation}\label{sec:solve eqs- power series}
We write $\hat{\gamma}(\theta,t)$, $\hat{u_r}(\theta,t)$, and $\hat{u}_{\theta}(\theta,t)$ as power series in the form of

\begin{equation}
    \label{Eq: taylor}
    \hat{\gamma}(\theta,\hat{t}) = \sum \limits_{n=0}^\infty  \hat{\gamma}_{n}(\theta)\hat{t}^n; \,\,\,\,
    \hat{u}_r(\theta,\hat{t}) = \sum \limits_{n=0}^\infty  \hat{u}_{r n}(\theta)\hat{t}^n, \,\,\,\,\,
    \hat{u}_{\theta}(\theta,\hat{t}) =\sum \limits_{n=0}^\infty \hat{u}_{\theta n}(\theta)\hat{t}^n.
\end{equation}

According to (\ref{Eq:xi_initial_eq}), for $\hat{\xi}$ and the factor ${1}/{\hat{\xi}}$ in (\ref{Eq:transport_dimensionless}), we have:
\begin{align}
    \label{Eq: xi_ur}\hat{\xi}&=1+\int \hat{u}_r\textrm{d}\hat{t}=1+\sum \limits_{n=0}^\infty \frac{1}{n+1} \hat{u}_{r n} \hat{t}^{n+1} ,\\
    \label{Eq: taylor_xi}\frac{1}{\hat{\xi}} &= \sum \limits_{n=0}^\infty  f_{n}(\theta)\hat{t}^n,
\end{align} 
where $f_n$s are Taylor series expansion coefficients of ${1}/{\hat{\xi}}$. For example, the first three coefficients, which are used in the final solution of this paper, can be shown to be: $f_{0}=1$, $f_{1}=-\hat{u}_{r 0}$ and $f_{2} =\hat{u}^2_{r 0} -\frac{1}{2} \hat{u}_{r 1}$.
We insert the power series (\ref{Eq: taylor}) and (\ref{Eq: taylor_xi}) into (\ref{Eq: u_r_dimensionless})-(\ref{Eq:transport_dimensionless})
and equating coefficients
we obtain
\begin{align} 
\label{Eq: u_r_dimensionless_power_series} 
\hat{u}_{rn}(\theta)&\approx\mathrm{p.v.}\; \frac{1}{4\pi} \int_{0}^{2\pi} \frac{\hat{\gamma}_n(\theta')}{\tan{\left[\left(\theta'-\theta\right)/2\right]}}\textrm{d}\theta'-\frac{1}{2\pi}\int\limits_0^{2\pi}\hat{\gamma}_n(\theta')\sin\left({\theta'-\theta}\right)\textrm{d}\theta',\\
    \label{Eq:u_c_dimensionless_power_series}
          \hat{u}_{\theta n}(\theta)&\approx\frac{1}{4\pi} \int_{0}^{2\pi}\hat{\gamma}_n(\theta')\textrm{d}\theta'+\frac{1}{2\pi}\int\limits_0^{2\pi}\hat{\gamma}_n(\theta')\cos\left({\theta'-\theta}\right)\textrm{d}\theta'- 
\begin{cases}
    \upchi,& \text{if } n=0\\
    0,              & \text{if } n>0
\end{cases}\\
    \label{Eq:transport_dimensionless_power_series}
     \hat{\gamma}_{n+1} &\approx  -\frac{1}{(n+1)} \left( \sum\limits_{j=0}^{n}f_{j} \sum\limits_{i=0}^{n-j} \frac{\partial (\hat{\gamma}_i \hat{u}_{\theta (n-j-i)})}{\partial \theta}\right)
\end{align}
The first term on the right-hand side of (\ref{Eq: u_r_dimensionless_power_series}) is a Cauchy principal value (p.v.) of an improper integral. The following identities are useful to solve this integral:
\begin{align}
    \mathrm{p.v.}\; \int_{0}^{2\pi} \frac{\sin{n x}}{\tan{\left[\left(x-b\right)/2\right]}}\textrm{d}x&=2\pi\cos{nb}, \label{eq:pv_int1} \\
    \mathrm{p.v.}\; \int_{0}^{2\pi} \frac{\cos mx}{\tan{\left[\left(x-b\right)/2\right]}}\textrm{d}x&=-2\pi\sin{mb}, \label{eq:pv_int2}
\end{align}
where $n\in\{1,2,...\}$, $m\in\{0,1,2,...\}$, and $b\in[0,2\pi]$. The complete derivation of these integrals can be found in the appendix \ref{appA}.

From (\ref{Eq:gamma_0}), $\hat{\gamma}_0(\theta)=\sin\theta+\upchi$. One can insert $\hat{\gamma}_0$ into (\ref{Eq: u_r_dimensionless_power_series}) and (\ref{Eq:u_c_dimensionless_power_series}) to respectively find $\hat{u}_{r0}(\theta)$ and $\hat{u}_{\theta 0}(\theta)$. Values of $\hat{u}_{r0}(\theta)$, $\hat{u}_{\theta 0}(\theta)$ and $f_0$ can be then inserted into (\ref{Eq:transport_dimensionless_power_series}) to find $\hat{\gamma}_1(\theta)$. This recursive process is repeated  until reaching the desired order of evaluation
for the power series of (\ref{Eq: taylor}). 
After $\hat{u}_{r}(\theta,\hat{t})$ is obtained using the developed recursive relations, the dimensionless shape of the vortex sheet $\hat{\xi}(\theta,\hat{t})$ is evaluated from (\ref{Eq: xi_ur}). The solutions for $\hat{\gamma}$, $\hat{u_r}$ and $\hat{u_{\theta}}$ up to $\mathcal{O}(\hat{t}^3)$ and for $\hat{\xi}$ up to $\mathcal{O}(\hat{t}^4)$ are written below

\begin{dmath}\label{eq:final_gamma}
 \hat{\gamma}(\theta,\hat{t}) = \sin (\theta ) + \upchi -\frac{1}{2} \hat{t} \sin (2 \theta ) +\hat{t}^2 \left(-\frac{1}{4} \upchi  \cos (2 \theta )+\frac{3}{16} \sin (3 \theta )-\frac{\sin (\theta )}{16}\right) +\hat{t}^3 \left(\frac{1}{12} \upchi ^2 \sin (2 \theta )-\frac{1}{48} \upchi  \cos (\theta )+\frac{5}{32} \upchi  \cos (3 \theta )+\frac{5}{96} \sin (2 \theta )-\frac{7}{96} \sin (4 \theta )\right),
\end{dmath}

\begin{dmath}\label{eq:u_r}
 \hat{u}_r(\theta,\hat{t}) = -\frac{1}{4} \hat{t} \cos (2 \theta )+ \hat{t}^2 \left(\frac{1}{8} \upchi  \sin (2 \theta )+\frac{3}{32} \cos (3 \theta )\right) + \hat{t}^3 \left(\frac{1}{24} \upchi ^2 \cos (2 \theta )-\frac{5}{64} \upchi  \sin (3 \theta )+\frac{5}{192} \cos (2 \theta )-\frac{7}{192} \cos (4 \theta )\right),
\end{dmath}

\begin{dmath}\label{eq:final_u_s}
 \hat{u}_{\theta}(\theta,\hat{t}) =
 \frac{1}{2}\sin (\theta ) - \frac{1 }{2}\upchi -\frac{1}{32} \hat{t}^2 \sin (\theta ) -\frac{1}{96} \hat{t}^3 \upchi  \cos (\theta ),
\end{dmath}

\begin{dmath}\label{eq:final_xi}
 \hat{\xi}(\theta,\hat{t}) = 1
 -\frac{1}{8} \hat{t}^2 \cos (2 \theta )+\hat{t}^3 \left(\frac{1}{24} \upchi  \sin (2 \theta )+\frac{1}{32} \cos (3 \theta )\right)+\hat{t}^4 \left(\frac{1}{96} \upchi ^2 \cos (2 \theta )-\frac{5}{256} \upchi  \sin (3 \theta )+\frac{5}{768} \cos (2 \theta )-\frac{7}{768} \cos (4 \theta )\right).
\end{dmath}
One can compute higher-order terms of (\ref{eq:final_gamma})-(\ref{eq:final_xi}), which may become relevant at increasing values of $\hat{t}$. However, 
since the above solution is developed based on the assumption that the vortex sheet remains approximately circular, increasing  deformation of the vortex sheet makes the solution inaccurate 
at large values of $\hat{t}$. 
For practical applications at large $\hat{t}$, in Appendix \ref{sec:appendix_empirical}  we propose an empirical formula that merges smoothly with the theoretical expression at small $\hat{t}$ (i.e., $|\hat{t}|\leq$2), while it has desired reasonable properties at large times (i.e., $|\hat{t}|>$2). 
In the next section, we prove the validity of the initial conditions in (\ref{Eq:gamma_0}) and (\ref{Eq:Gamma_r}). Moreover, values of $\xi_0$, $\gamma_b$ and $\gamma_r$ are determined as functions of turbine operating conditions.

\subsection{Setting vortex sheet initial conditions at yawed turbine location}\label{subsect:initial_vorticity}
In this section, we determine the vorticity shedding from a yawed rotor disk (i.e., $\gamma_0(\theta)=\gamma(\theta,0)$). According to Kutta-Jokowski theorem, lift force is proportional to the amount of circulation around a lifting airfoil. This means that an airfoil can be replaced with a \emph{bound} vortex. 
Also, 
for any airfoil with finite span, \emph{free} vortices must trail downstream from both sides of the bound vortex to infinity, forming a \emph{horseshoe vortex} \citep{Anderson2011Fundamentals}. 
Turbine blades rotate and produce power due to their generated lift force, and 
vorticity is shed from the root and the tip of rotor blades. In addition, the whole yawed rotor can be assumed as a big finite-span airfoil with the lateral component of the thrust force regarded as the lift force. Therefore, in order to find the total shedding vorticity at the rotor disk, we need to determine those due to both yaw offset and rotating blades. In the following, we assume that the yawed rotor 
can be modelled as a rotating actuator disk.

\subsubsection{Vorticity shedding due to turbine yaw}\label{sec:initial_yaw_offset}
\begin{figure}
\centerline{\includegraphics[width=\textwidth]{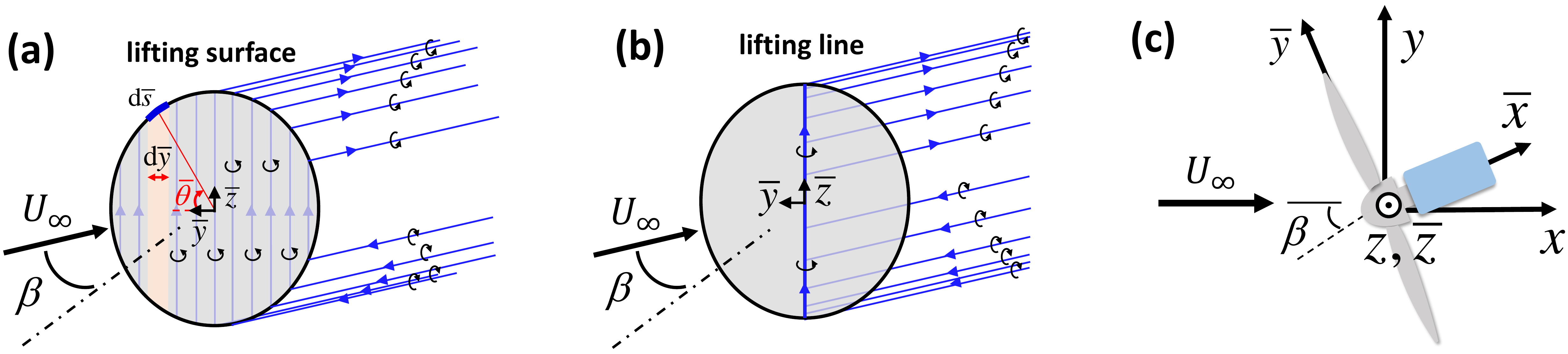}}
\caption{Vorticity shedding from a yawed actuator disk, modelled (a) as a lifting surface, and (b) as a lifting line. (c) A schematic of different coordinate systems used in this paper. }\label{fig:initial_yaw}
\end{figure}
Prior studies have suggested two different approaches to model the distribution of circulation at a yawed disk. By modeling a yawed disk as a \emph{lifting line}, the circulation is concentrated on a vertical line at the centre of the rotor with an elliptical distribution spanning from the bottom tip to the top tip of the rotor \citep{shapiro2018}. Alternatively, vorticity due to yaw offset can be assumed to shed from the circumference of the rotor \citep{zong2020}. \citet{Martinez2020equiva} used vortex cylinder theory to state the equivalency of these two methods. \citet{shapiro2020decay} proved that both vorticity distributions yield the same induced velocity inside the radius of the rotor disk. To determine the reference circulation density $\gamma_b$ needed in  (\ref{Eq:gamma_0}) and to provide more physical insight, we build upon the literature to show that the equivalency of these two vorticity distributions can be also verified simply by rearranging the position of horseshoe vortices over a yawed disk.

 Figure \ref{fig:initial_yaw}a shows a schematic of a yawed actuator disk modelled as a \emph{lifting surface} with a constant vortex strength of $\gamma_b$ in $\bar{z}$-direction. The coordinate system $(\bar{x},\bar{y},\bar{z})$ is defined based on the rotor plane as shown in figure \ref{fig:initial_yaw}c, and its respective polar coordinate system ($\bar{r},\bar{\theta}$) is defined such that $\bar{r}=\sqrt{\bar{y}^2+\bar{z}^2}$ and $\tan\bar{\theta}=\bar{z}/\bar{y}$. The lifting surface shown in figure \ref{fig:initial_yaw}a can be envisaged as a surface with an infinite number of horseshoe vortices uniformly distributed across the yawed disk. 
 The bound circulation at a given vertical position $z$ is given by
 \begin{equation}
     \Gamma(z)=\int\limits_{-\sqrt{R^2-\bar{z}^2}}^{\sqrt{R^2-\bar{z}^2}}\gamma_b\mathrm{d}\bar{y}=2\gamma_b\sqrt{R^2-\bar{z}^2},
     \label{Eq:Gamma_bound}
 \end{equation}
  where $R$ is the rotor radius. Note that according to (\ref{Eq:Gamma_bound}), the vertical distribution of the bound circulation for the lifting surface with a constant vortex strength $\gamma_b$ is elliptical.  This means that if we concentrate all these horseshoe vortices on a vertical line at the centre of the disk, the lifting surface is transformed to a lifting line with an elliptical distribution of circulation, like the one used by \citet{shapiro2018}, as shown in figure \ref{fig:initial_yaw}b. In this case, trailing vortices shed from all along the lifting line, because it consists of horseshoe vortices that vary in size. From (\ref{Eq:Gamma_bound}), the maximum value of bound circulation for the lifting line, denoted by $\Gamma_b$, occurs at at $\bar{z}=0$, and its value is equal to $2R\gamma_b$. From the lifting line theory, we know that $\Gamma_b=-U_{h} C_T R \cos^2{\beta}\sin{\beta}$ \citep{shapiro2018}, where $C_T$ is the turbine thrust coefficient. The value of $C_T$ is given by  
\begin{equation}\label{eq:thrust_coef}
    C_T= \frac{2T}{\rho \pi R^2 U_{h}^2 \cos^2\beta},
\end{equation}
where $T$ is the total magnitude of the turbine thrust force, $\rho$ is the air density and $U_{h}$ is the inflow velocity at the hub height. Note that this definition is the same as the one used in \citet{shapiro2018}, but it is different from the one used in some other prior studies \citep[e.g.,][]{Burton1995,bastankhah2016experimental}. Since $\gamma_b=\Gamma_b/2R$, the value of $\gamma_b$ is given by
\begin{equation}
    \gamma_b=-\frac{1}{2}U_{h} C_T \cos^2{\beta}\sin{\beta}.
    \label{Eq:gamma_b_final}
\end{equation}

Next, we determine the distribution of trailing vortices shedding from the circumference of the lifting surface. The circulation of the vortex shedding from an infinitesimal circumferential element $\mathrm{d}s$, where $\mathrm{d}\bar{s}=R\mathrm{d}\bar{\theta}$, is  $\mathrm{d}\Gamma_{\rm shed, yaw}=\gamma_b \mathrm{d}\bar{y}$. Given that $\mathrm{d}\bar{y}=\mathrm{d}\bar{s}\sin{\bar{\theta}}$, we obtain
\begin{equation}
    \mathrm{d}\Gamma_{\rm shed, yaw}=\gamma_b\sin{\bar{\theta}}\mathrm{d}\bar{s},
    \label{Eq:Gamma_shed_surface}
\end{equation}
For the lifting line, on the other hand, the magnitude of circulation of trailing vortex over the segment $\mathrm{d}\bar{z}$ is equal to -$\left(\mathrm{d}\Gamma/\mathrm{d}\bar{z}\right)\mathrm{d}\bar{z}$ \citep{Anderson2011Fundamentals}. From (\ref{Eq:Gamma_bound}) and $\Gamma_b=2R\gamma_b$, we obtain 
\begin{equation}
    \textrm{d}\Gamma_{\rm shed, yaw}=\frac{\Gamma_b \bar{z} \mathrm{d}\bar{z}}{R\sqrt{R^2-\bar{z}^2}}.
    \label{Eq:Gamma_shed_line}
\end{equation}
Using the variable change $\bar{z}=R\sin{\bar{\theta}}$, one can easily show that $\mathrm{d}\Gamma_{\rm shed}$ for the lifting surface at any $\bar{\theta}$ (\ref{Eq:Gamma_shed_surface}) is half of the one of the lifting line (\ref{Eq:Gamma_shed_line}) at the respective height $\bar{z}$. Note that for the lifting surface, at a given height, trailing vortices shed at both angles of $\bar{\theta}$ and $(\pi - \bar{\theta})$ with the same magnitude of circulation. Therefore, trailing vortices shedding from the lifting line and the lifting surface vary with height in a similar manner. \black It is also worth mentioning that the results presented here are in agreement with those obtained from the skewed vortex cylinder theory \citep{coleman1945,branlard2016}.  By modelling a yawed turbine wake as a skewed vortex cylinder, \citet{Martinez2020equiva} stated that the dominant vorticity shedding from the rotor is the tangential vorticity vector, which lies in the rotor plane. The streamwise projection of this tangential vorticity is equal to the one found in the present work \eqref{Eq:Gamma_shed_surface} (c.f., equation 9 in \citet{Martinez2020equiva}). \black

\subsubsection{Inclusion of wake angular momentum effects}\label{sec:initial_rotation}
\begin{figure}
\centerline{\includegraphics[width=.8\textwidth]{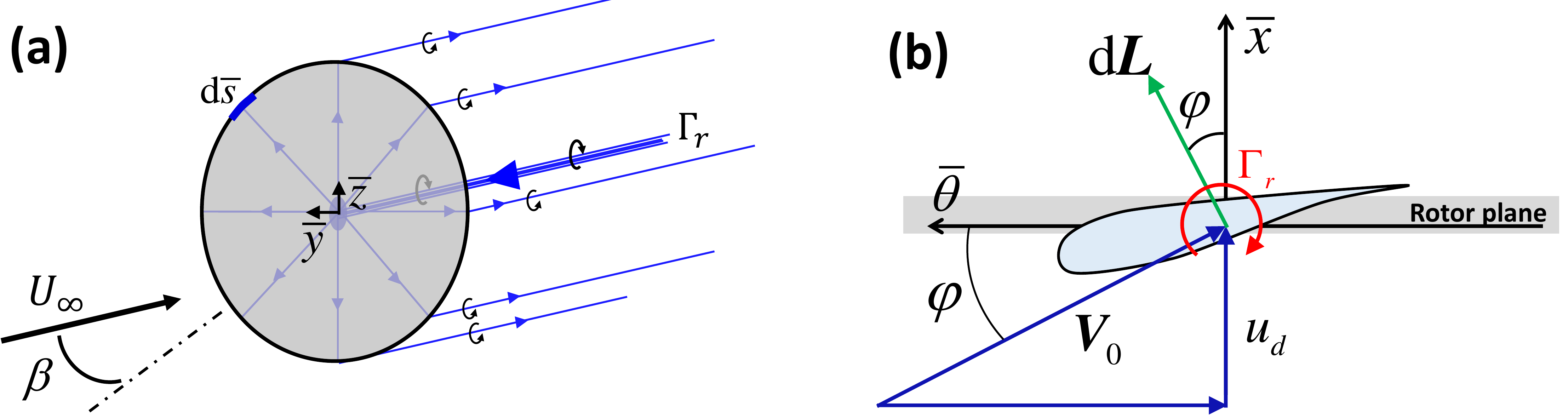}}
\caption{(a) Modelling a turbine rotor as a rotating actuator disk. (b) The velocity triangle for a rotor blade element.}\label{fig:initial_rotation}
\end{figure}

In this section, we determine the value of $\gamma_r$ required in (\ref{Eq:gamma_0}) and (\ref{Eq:Gamma_r}). 
Based on the  method of Joukowsky that models a turbine blade as one single horse-shoe vortex with constant bound circulation (see \citet{okulov2012betz} for historical background), two free trailing vortices with the same magnitude of circulation are shed from both root and tip ends of each turbine blade. Under the assumption of large number of blades, this creates a vortex system consisting of a bound vortex disk, an axial root vortex and a tubular vortex sheet as shown in the figure \ref{fig:initial_rotation}a. 
Let us denote the circulation of the root trailing vortex with $\Gamma_r$. As the amount of circulation along any horseshoe vortice remains constant, the bound circulation on the rotor disk at any radial position should be the same as $\Gamma_r$. According to the Kutta–Joukowsky theorem, the bound circulation over an annular ring at a radial position $\bar{r}$ and thickness of $\textrm{d}\bar{r}$ on the rotor disk generates a lift force $\textrm{d}\mathbf{L}$, which amounts to \citep{okulov2010maximum}
\begin{equation}\label{eq:kutta-joukowsky}
    \textrm{d}\mathbf{L}=\rho \mathbf{V_0}\times \mathbf{e_{\bar{r}}}\Gamma_r\textrm{d}\bar{r},
\end{equation}
where $\mathbf{V_0}$ is the resultant relative wind velocity experienced by the blade element as shown in figure \ref{fig:initial_rotation}b. In this figure, $\phi$ denotes the angle between $\mathbf{V_0}$ and $\bar{\theta}$-direction, and $u_d$ is the component of $\mathbf{V_0}$ in the $\bar{x}$-direction. The tangential component of $\textrm{d}\mathbf{L}$ produces power $P$, which is given by
\begin{equation}\label{eq:dP:kutta-Jowk}
    \textrm{d}P=\Omega\textrm{d}Q=\rho\Omega\Gamma_r u_d \bar{r}\textrm{d}\bar{r},
\end{equation}
where $\Omega$ is the turbine rotational velocity, and $\textrm{d}Q$ is the torque generated by the given annular ring. From the axial momentum theory, the power generated by the annular ring can be written as the product of $u_d$ and $\textrm{d}T$, where $\textrm{d}T$ is the thrust force exerted on the annular ring. Therefore, we obtain an additional equation for $\textrm{d}P$ as follows 
\begin{equation}\label{eq:dP:momentum_eq}
    \textrm{d}P=u_d\textrm{d}T=\rho u_d U_{\infty}^2C_T\cos^2\beta(\pi\bar{r})\textrm{d}\bar{r}.
\end{equation}
Note that to derive (\ref{eq:dP:momentum_eq}), the local thrust coefficient for a given annular ring is assumed to be the same as its value for the whole rotor defined in (\ref{eq:thrust_coef}). This is a correct assumption for the Joukowsky vortex model  \citep{van_Kuick2015rotor}. Equating (\ref{eq:dP:kutta-Jowk}) and (\ref{eq:dP:momentum_eq}) leads to
\begin{equation}\label{eq:Gamma_r_final value}
  \Gamma_r=\frac{\pi R}{\lambda} U_{h}C_T\cos^2\beta,  
\end{equation}
where $\lambda$ is the tip-speed ratio and defined as $\Omega R/U_{h}$.

As seen in figure \ref{fig:initial_rotation}a, the trailing vorticity sheds over the circumference of the rotor disk. It is evident that the value of circulation for the vorticity shedding over the circumferential element $\mathrm{d}\bar{s}$ is given by
\begin{equation}\label{eq:dGamma_shed_rotation}
    \textrm{d}\Gamma_{\rm shed, rot}=\gamma_r\textrm{d}\bar{s},
\end{equation}
where $\gamma_r=-\Gamma_r/(2\pi R)$. The variable $\gamma_r$ denotes the strength of the shedding tubular vortex sheet, and from (\ref{eq:Gamma_r_final value}), 
\begin{equation}\label{eq:gamma_r_final value}
    \gamma_r=-\frac{1}{2\lambda} U_{h}C_T\cos^2\beta.
\end{equation}

It is important to note that as discussed by \citet{branlard2016}, the above-mentioned shedding vortices are actually in the direction of the wake centreline axis that forms an angle with the streamwise coordinate $x$. Prior studies \citep[e.g.,][]{coleman1945} however showed that the angle between the wake centreline and the $x-$coordinate is expected to be much smaller than the turbine yaw angle. Therefore, we assume that both axial root and tip shedding vortices are in the $x$-direction for simplicity. 

\subsubsection{Total vorticity shedding from a yawed turbine}
From findings of \S\ref{sec:initial_yaw_offset} and \S\ref{sec:initial_rotation}, we can determine the total vorticity shedding from a yawed rotor due to both yaw offset and rotating blades as a function of turbine operating conditions. From (\ref{Eq:Gamma_shed_surface}) and (\ref{eq:dGamma_shed_rotation}), the value of the dimensionless initial vortex strength $\hat{\gamma}_0$ in the $(r,\theta)$ polar coordinate system is given by
\begin{equation}\label{eq:totol_initial_vorticity}
    \hat{\gamma}_0(\theta)=\frac{\gamma_0}{\gamma_b}=\sin\theta+\upchi \black\quad\quad\textrm{at}\; r=\xi_0,
\end{equation}
where from (\ref{Eq:gamma_b_final}) and (\ref{eq:gamma_r_final value}),
\begin{equation}\label{eq:rotation_rate}
    \upchi=\frac{\gamma_r}{\gamma_b}=\frac{1}{\lambda\sin\beta}.
\end{equation}
The variable $\upchi$ called the rotation rate is the ratio of the strength of vortex generation due to rotating blades to the one generated due to yaw offset. For the limiting cases of $\lambda=\pm\infty$, $\upchi$ is equal to zero, and the shedding vorticity is only due to the yaw offset. As the tip-speed ratio $\lambda$ goes to infinity, the amount of torque generated by the turbine goes to zero. Therefore, according to the conservation of angular momentum, there should be no wake rotation downwind of the actuator disk in the limiting cases of $\lambda=\pm\infty$. Hereafter, the term \emph{non-rotating} wake refers to the wake of an actuator disk with an infinite tip-speed ratio $\lambda$, while \emph{rotating} wake refers to the wake of an actuator disk with a finite value of $\lambda$. 

\subsubsection{Initial shape of the vortex sheet}
The vortex sheet sheds from the circumference of the rotor, so it initially has a shape similar to the projected frontal area of the yawed disk, which is an ellipse with a semi-major axis of $R$ in the $z$-direction, and a semi-minor axis of $R\cos\beta$ in the $y-$direction. \black The disk-averaged velocity normal to the rotor $u_d$ is equal to $U_{\infty}\cos\beta(1-a)$, where $a$ is the turbine induction factor, and it is given by $0.5(1-\sqrt{1-C_T\cos^2\beta})$ \citep{Burton1995}. 
\black
Behind the turbine, the rotor streamtube area expands further as 
pressure recovers to the background value \citep{manwell2010wind}. \black At this location, the streamwise velocity is given by $U_{\infty}(1-2a)$ \citep{bastankhah2016experimental,shapiro2018}. From continuity, \black $A_*$, the ratio of the expanded streamtube area to the projected frontal area of the rotor is therefore given by
\begin{equation}\label{eq:A_*}
    A_*=\frac{\left(1-a\right)\cos\beta}{1-2a}\frac{1}{\cos\beta}=\frac{1+\sqrt{1-C_T\cos^2\beta}}{2\sqrt{1-C_T\cos^2\beta}},
\end{equation}
Neglecting the distance between the rotor and the end of the streamtube expansion, 
we set the initial wake area  enclosed by the vortex sheet at $t=0$ to be the projected frontal area of the rotor times $A_*$. Therefore, $\xi_0(\theta)$ has an elliptical shape expressed by
\begin{equation}\label{Eq:xi_0}
    \xi_{0}(\theta)=R\sqrt{A_*}\frac{|\cos{\beta}|}{\sqrt{1-\sin^2{\beta}\sin^2{\theta}}}.
\end{equation}
For a small yaw angle $\beta$, the vortex sheet initially has an approximately circular shape. Therefore, one can approximate $\xi_0$ with $\tilde{\xi}_0$ given by
\begin{equation}\label{Eq:xi_0-circular}
    \tilde{\xi}_0\approx R\sqrt{A_*}.
\end{equation}

\subsection{Vortex sheet lateral deflection}\label{sec:lateral_deflection}
The analytical  solutions of the vortex sheet shape developed earlier are represented in the $(r,\theta)$ polar coordinate system, which is attached to the vortex sheet centre $C$. Therefore, in order to fully determine the locus of the vortex sheet with respect to a stationary coordinate system, we also need to compute how $y_c$ and $z_c$ vary with time or downstream distance (i.e., wake deflection). From figure \ref{fig:schematic_vortex}c, the Kutta-Joukowski theorem can be used to obtain 
\begin{equation}\label{Eq:y_c_initial_eq}
\hat{y}_c=\frac{1}{2\pi}\int\limits_0^{\hat{t}}\int\limits_0^{2\pi}\hat{\gamma}(\theta',\hat{t})\sin\theta'\textrm{d}\theta'\textrm{d}\hat{t},
\end{equation}
\begin{equation}\label{Eq:z_c_initial_eq}
\hat{z}_c=\frac{-1}{2\pi}\int\limits_0^{\hat{t}}\int\limits_0^{2\pi}\hat{\gamma}(\theta',\hat{t})\cos\theta'\textrm{d}\theta'\textrm{d}\hat{t}.
\end{equation}
where $\hat{y}_c=y_c/\xi_0$ and $\hat{z}_c=z_c/\xi_0$. Inserting $\hat{\gamma}(\theta,\hat{t})$ from (\ref{eq:final_gamma}) into (\ref{Eq:y_c_initial_eq}) and (\ref{Eq:z_c_initial_eq}) and performing the integration lead to
\begin{equation}\label{Eq:y_c_eq2}
  \hat{y}_c=\frac{\hat{t}}{2}-\frac{\hat{t}^3}{96}, 
\end{equation}
\begin{equation}\label{Eq:z_c_eq2}
  \hat{z}_c=\frac{\upchi\hat{t}^4}{384}.
\end{equation}
From (\ref{Eq:z_c_eq2}), the vertical displacement of $C$ is zero when $\upchi=0$ (i.e., actuator disks with non-rotating wake) as expected from symmetry and consistent with prior experimental and numerical works \citep[e.g.,][]{Howland2016,bastankhah2016experimental,Bartl2018}. Comparison of (\ref{Eq:y_c_eq2}) and (\ref{Eq:z_c_eq2}) for nonzero values of $\upchi$ shows that $\hat{z}_c$ is nonzero but still considerably smaller than $\hat{y}_c$ for small values of $\hat{t}$. Therefore, we neglect $\hat{z}_c$ in this work for simplicity.

\begin{figure}
\begin{center}
 
\begin{subfigure}{0.45\textwidth}
\includegraphics[width=0.9\linewidth]{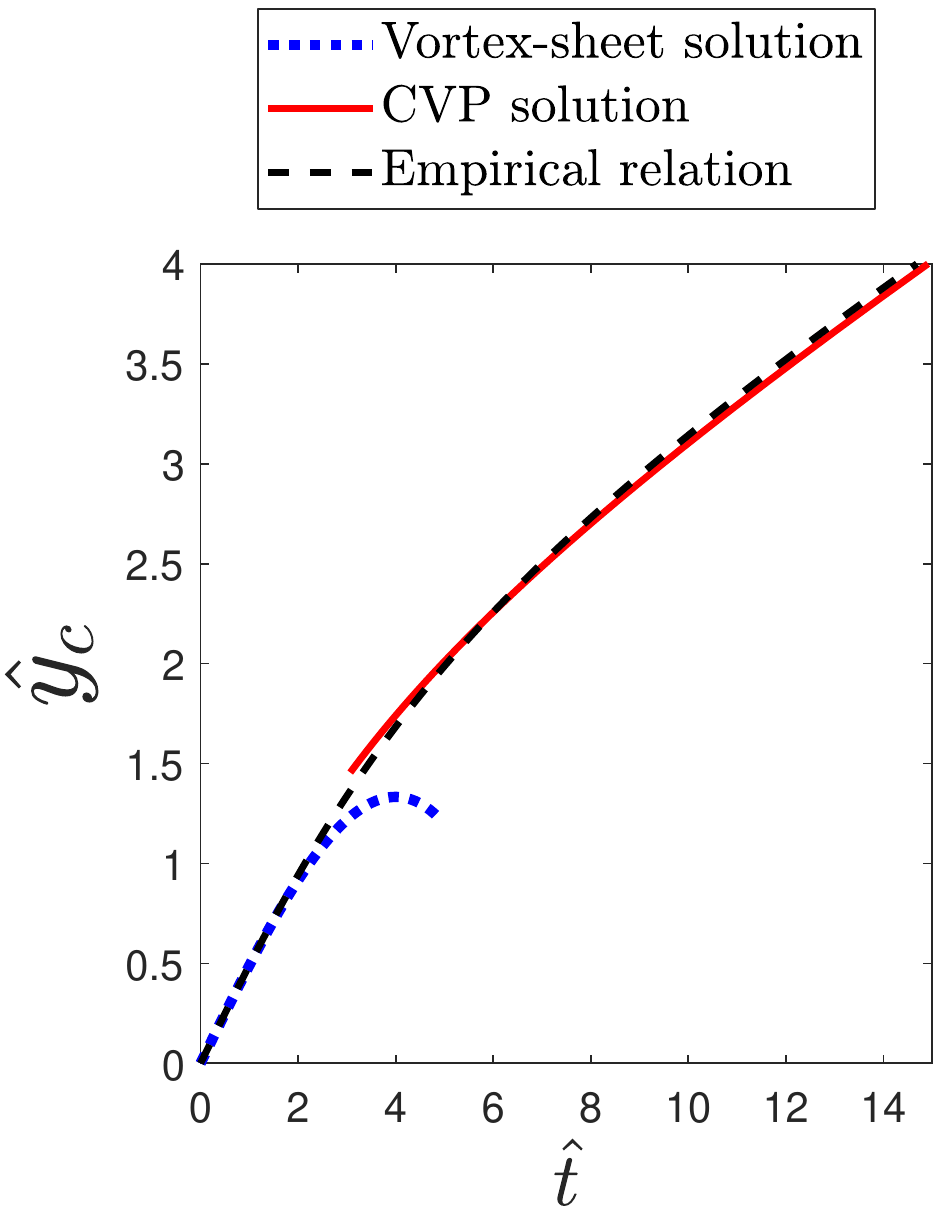} 
\caption{}
\end{subfigure}
\begin{subfigure}{0.3\textwidth}
\includegraphics[width=0.9\linewidth]{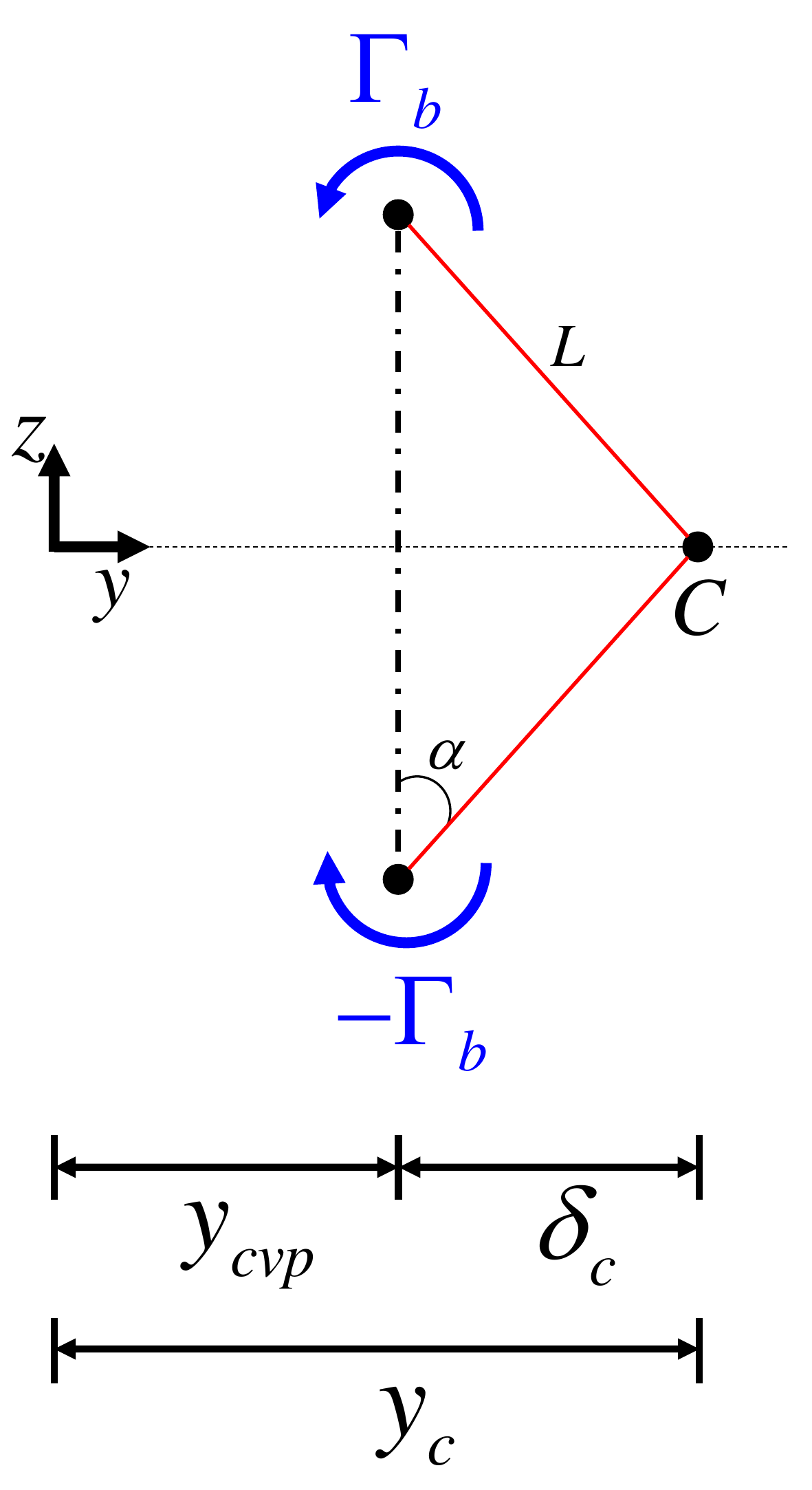}
\caption{}
\end{subfigure}
\caption{(a) Lateral deflection of the vortex sheet centre $y_c$ based on modelling the shed vorticity either as an approximately circular vortex sheet (\ref{Eq:y_c_eq2}) or a CVP (\ref{Eq:y_c_cvp}). The empirical relation (\ref{Eq:empirical}) provides predictions similar to the former approach at small $\hat{t}$, while it tends to the latter solution at large times. (b) Schematic of modelling the CVP shedding from a yawed rotor. }
\label{fig:y_c}

\end{center}
\end{figure}

It is worth remembering that to derive the analytical solution for the deformation of the vortex sheet, we assumed that the shape of the vortex sheet does not largely deviate from a circle. Although this is an acceptable assumption for small values of yaw angle and time, it is less accurate for large values of times, when the vortex sheet rolls up and forms a CVP. As shown in figure \ref{fig:y_c}a, based on (\ref{Eq:y_c_eq2}), the value of $\hat{y}_c$ may even decrease with an increase of $\hat{t}$, which is clearly unphysical. Since we expect that at large times (or downstream distances) a CVP can  more realistically represent the vorticity shedding from the yawed rotor, we enhance our model for $\hat{y}_c$ so that at large distances it tends to the situation of a CVP instead of using the truncated series vortex sheet solution. 

 Figure \ref{fig:y_c}b shows a schematic of a CVP. The CVP has the circulation of $\Gamma_b$ \citep{shapiro2018} as shown in figure \ref{fig:y_c}b. The lateral position of the CVP is denoted by $y_{cvp}$, and the lateral distance between the wake centre $C$ and the CVP is denoted by $\delta_c$ in the figure. The vertical spacing between counter-rotating vortices is equal to $2\xi_0$ and it is assumed to remain constant. Initial values of $y_c$, $y_{cvp}$ and $\delta_c$ are zero.  In this analysis, only the vorticity shed  due to the yaw offset is  considered as the effect of shed vorticity due to rotating blades on the wake deflection is expected to be small.  Our objective is to find the variation of $y_c$ with time (or downstream distance), but let us first determine how $\delta_c$ varies with time. According to the Biot-Savart law:
\begin{equation}
    v_c=\frac{2\Gamma_b}{2\pi L}\cos\alpha=\frac{\Gamma_b \xi_0}{\pi \left(\xi_0^2+\delta_c^2\right)},
\end{equation}
where $v_c$ is the lateral velocity of $C$ induced by the CVP, and $L$ and $\alpha$ are defined in figure \ref{fig:y_c}b. The CVP also moves with a lateral velocity of $v_{cvp}=\Gamma_b/(4\pi\xi_0)$ due to its self-induced velocity. Therefore, one can write
\begin{equation}\label{eq:d detal/d t}
    \frac{\textrm{d}\delta_c}{\textrm{d}t}=v_c-v_{cvp}=\frac{\Gamma_b \xi_0}{\pi \left(\xi_0^2+\delta_c^2\right)}-\frac{\Gamma_b}{4\pi\xi_0}=\frac{\Gamma_b}{4\pi \xi_0}\left(\frac{3\xi_0^2-\delta_c^2}{\xi_0^2+\delta_c^2}\right)
\end{equation}
It is interesting to note that according to (\ref{eq:d detal/d t})  $\delta_c$ increases until $\delta_c$ approaches $\sqrt{3}\xi_0$. At this time $v_c=v_{cvp}$ and therefore the relative position of C with respect to the CVP does not change anymore, and $\delta_c$ remains equal to $\sqrt{3}\xi_0$ afterwards. 

Next, we approximate $\Gamma_b \approx 2\xi_0\gamma_b$ in (\ref{eq:d detal/d t}). We then integrate (\ref{eq:d detal/d t}) (with separation of variables, $\delta_c$ and $t$) and write the solution in the dimensionless form. This yields an implicit expression for $\hat{\delta}_c(\hat{t})$ (it will be later expressed explicitly using an empirical formula):
\begin{equation}\label{eq:delta_c}
    \frac{\hat{t}}{2\pi}=\frac{-2}{\sqrt{3}}\ln{\left(\frac{\sqrt{3}-\hat{\delta}_c}{\sqrt{3}+\hat{\delta}_c}\right)}-\hat{\delta}_c,
\end{equation}
where $\hat{\delta}_c=\delta_c/\xi_0$. Note that to derive (\ref{eq:delta_c}), we assume $\hat{\delta}_c < \sqrt{3}$. Given that $\hat{y}_c=\hat{\delta}_c+\hat{y}_{cvp}$, we have
\begin{equation}\label{Eq:y_c_cvp}
    \hat{y}_c=\hat{\delta}_c +\frac{\hat{t}}{2\pi}. 
\end{equation}
Predictions of $\hat{y}_c$ based on modelling the vortex sheet as a CVP using (\ref{eq:delta_c}) and (\ref{Eq:y_c_cvp}) are shown in figure \ref{fig:y_c}a. As discussed earlier, the solution for $\hat{y}_c$ based on the CVP is expected to provide acceptable predictions at large values of $\hat{t}$, while the solution based on an approximately circular vortex sheet (\ref{Eq:y_c_eq2}) works better at short values of $\hat{t}$. An empirical formula that merges these two behaviors to leading order in both limits can be written as    
\begin{equation}\label{eq:y_c_empirical_generic}
    \hat{y}_c=\frac{c_1|\hat{t}|^3+c_2\hat{t}^2+c_3|\hat{t}|}{c_4\hat{t}^2+c_5|\hat{t}|+c_6}\textrm{sgn}(\hat{t}),
\end{equation}
where $\textrm{sgn}(x)$ is the sign function of $x$, and $c_1,...,c_6$ are polynomial coefficients, which need to be determined.  Note that, similar to the vortex sheet and CVP solutions, the empirical relation is an odd function, so the wake deflection is opposite for turbines with opposite yaw angles. To find suitable values of polynomial coefficients ($c_1,...,c_6$), we match the series expansion of (\ref{eq:y_c_empirical_generic}) at $\hat{t}\to 0$ and $\hat{t}\to \infty$ with $\hat{t}/2-\hat{t}^3/96$ and $\hat{t}/(2\pi)+\sqrt{3}$, respectively. This leads to a system of equations that needs to be solved. So we obtain
\begin{equation}\label{Eq:empirical}
\hat{y}_c=\frac{(\pi-1)|\hat{t}|^3+2\sqrt{3}\pi^2\hat{t}^2+48(\pi-1)^2|\hat{t}|}{2\pi(\pi-1)\hat{t}^2+4\sqrt{3}\pi^2|\hat{t}|+96(\pi-1)^2}\textrm{sgn}(\hat{t}).
\end{equation}
Equation (\ref{Eq:empirical}) provides predictions similar to (\ref{Eq:y_c_eq2}) at small values of $\hat{t}$ and approaches (\ref{Eq:y_c_cvp}) at large values of $\hat{t}$ as shown in figure \ref{fig:y_c}a.

\subsection{Comparison with numerical simulations} \label{sec:results_laminar}
In this section, we compare predictions of the vortex sheet (i.e., wake edge) shape $\xi(\theta,t)$ based on the new proposed model with numerical simulation data. For simulations,  the pseudo-spectral large-eddy simulation (LES) code LESGO is applied. LESGO has been used in prior works \citep{Calaf2010a, Stevens2014a, Verhulst2015a, Stevens2018a, Martinez2018a, shapiro2018} to simulate flow past wind turbines and wind farms.  It has been validated by detailed comparisons with several other LES codes \citep{Martinez2018a}. \black Turbines are simulated using the actuator disk model with rotation (ADM-R)\black. See the Appendix \ref{AppC:LES} for more information about the LESGO code and the LES setup of this study. Under uniform inflow conditions the role of turbulence is minimal, and the code runs mostly as an inviscid solver with regularization, as it was also used in \cite{shapiro2018}.  Simulations are performed for a range of local thrust coefficients $C_T'= 0.80, 1.0,$ and $1.33$, yaw angles $\beta = 10^\circ$, $20^\circ$, and $30^\circ$, and rotation rates $\upchi = 0$, 0.25, and 0.5. Note that according to (\ref{Eq:gamma_b_final}) this means that $\gamma_b$ and $\hat{t}$ are negative and the curling is expected to be in the opposite direction of that shown in the sketch in figure \ref{fig:schematic_vortex}, i.e. in the LES the wake is being deflected in the negative $y$-direction. Also, it is worth remembering that the rotation rate $\upchi$ depends on both yaw angle $\beta$ and tip speed ratio $\lambda$. \black According to \eqref{eq:rotation_rate}, for a utility-scale wind turbine with a tip speed ratio $\lambda=8$ and yaw angle $\beta=15^{\circ}-30^{\circ}$, rotation rate $\upchi$ varies between $0.25-0.5$. The non-rotating case commonly used in the LES corresponds to an infinite tip speed ratio and reduces to the standards actuator disk model (ADM) without rotation. \black 
The local thrust coefficient $C_T'$ is related to the thrust coefficient $C_T$ through $C_T=16C_T'/(4+C_T'\cos^2\beta)^2$ \citep{shapiro2018}. A fringe forcing region is used to force the flow back to laminar inflow when using periodic boundary conditions in the $x$-direction. Excluding this fringe region the effective domain has sides that are \black$L_x = 15.12D$\black, $L_y = 5.76D$, and $L_z = 5.76D$ long. A uniform grid with $N_x = 384$ effective grid points in the streamwise direction and $N_y = N_z = 192$ grid points in the spanwise and vertical directions are used. The centre of the actuator disk is placed 3.6D from the inlet of the domain. 



In order to determine the shape of the wake edge based on the developed model, we need to first compute the value of $\hat{t}=\gamma_bt/\xi_0$ where $\xi_0$ can be approximated with $\tilde{\xi}_0$ (\ref{Eq:xi_0-circular}) and $t=x/U_{con}$. \black Although the convection velocity $U_{con}$ in turbine wakes changes with the streamwise distance, it is approximated with a constant value in this study, as done in prior studies \citep[e.g.,][]{shapiro2020decay}. \black
 For cases with no incoming turbulence, the turbine wake does not significantly interact with the surrounding flow, and it experiences a slow recovery. In this case, the streamwise velocity profile in the central part of the wake can be modelled as a top-hat core (i.e., potential core) with a constant velocity $U_0$ equal to $U_{\rm{in}}\sqrt{1-C_T\cos^2\beta}$ \citep{bastankhah2016experimental, shapiro2018}, where $U_{\rm{in}}$ is the incoming velocity. The top-hat core is surrounded by a shear layer in which the velocity changes from $U_0$ to $U_{\rm{in}}$. Therefore, we approximate the convection velocity 
with $U_{con}=0.5(U_0+U_{\rm{in}})$.
For instance, based on this definition, $|\hat{t}|=2$ (i.e., limiting value for using the analytical model) corresponds to a streamwise distance in the range of  $12R-29R$ for a turbine with $C_T'=1.33$ (i.e., $C_T\approx 0.75$ for $\beta=0$), and $\beta=30^{\circ}-10^{\circ}$. 

\begin{figure}
\centerline{\includegraphics[width=.9\textwidth]{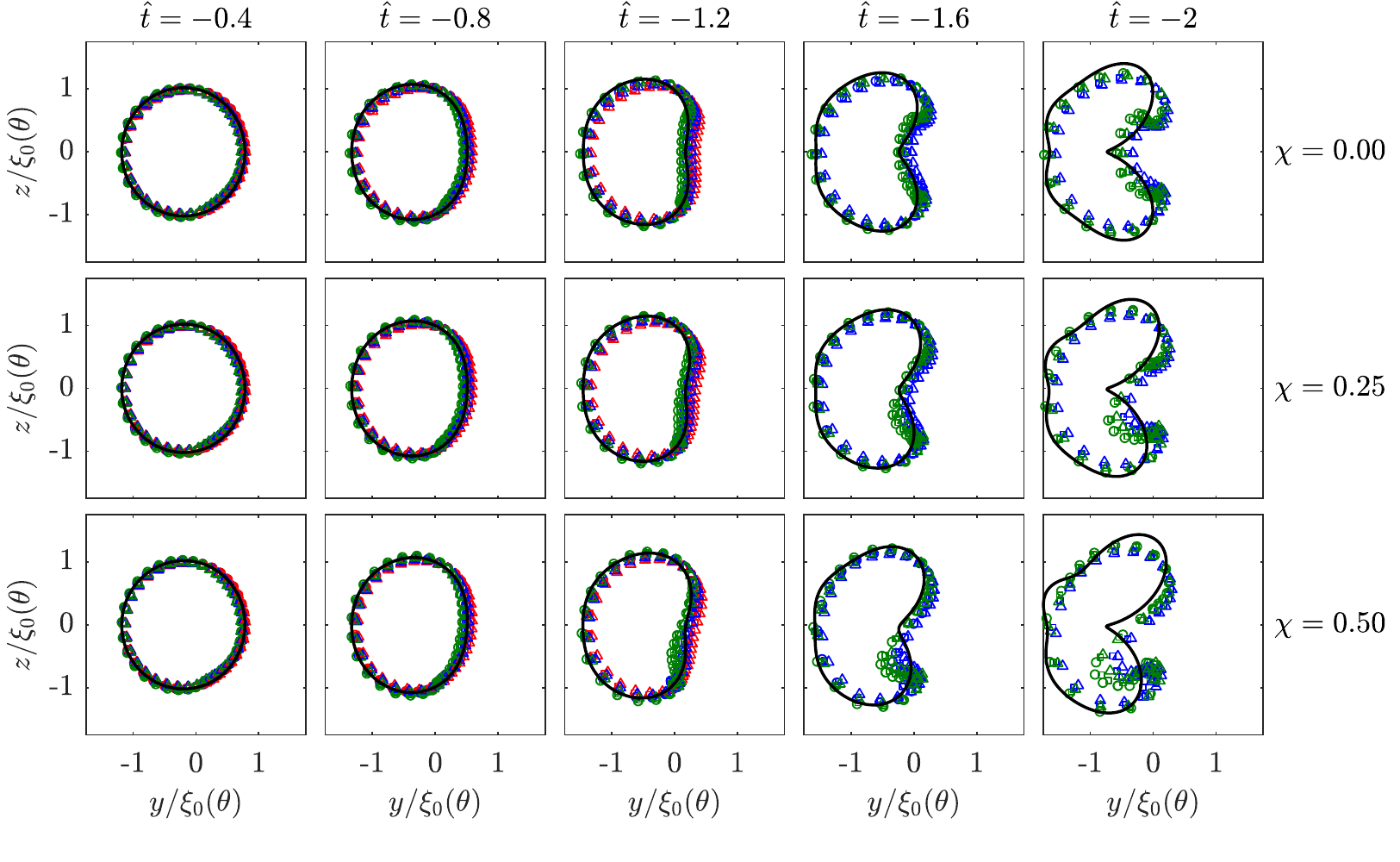}}
\caption{Dimensionless shape of the wake of yawed wind turbines in uniform inflow for $C_T' = 0.8$ ($\circ$), $C_T' = 1$ ($\square$), and $C_T' = 1.33$ ($\triangle$) and yaw angles $\beta=10^\circ$ (red),  $\beta=20^\circ$ (blue), $\beta=30^\circ$ (green) at various evolution times $\hat{t}$ and rotation rates $\upchi$. The analytical model (black solid line) 
is shown for comparison. \black Note that results of $\beta=10^{\circ}$ are not shown for $\hat{t}=-1.6$ and $-2.0$ because for this yaw angle they correspond to downwind distances that exceed the computational domain. \black}
\label{fig:uniform-that}
\end{figure}

The analysis presented in \S\ref{sec:model_derivation} suggests that the wake shape, non-dimensionalised by $\xi_0(\theta)$, only depends on the dimensionless time $\hat{t}$, and the rotation rate $\upchi=1/\lambda\sin\beta$. As a first test of the model 
we compare the dimensionless model predictions with LES results normalized such that they can be presented as function of 
$\hat{t}$ and $\upchi$. For the LES data, the edges of the wake are identified by tracking the edge of the streamtube that passes through the face of the actuator disk which is appropriate in this case due to the lack of turbulent mixing.  Results are shown in figure~\ref{fig:uniform-that}, which shows that the LES data for turbines with different operating conditions approximately collapse onto the same wake profile curve for given values of $\hat{t}$ and $\upchi$, in agreement with the proposed theory. 
The figure also shows that the proposed analytical model is able to capture the scaled wake shape. 
At larger time magnitudes 
and large rotation rates 
some discrepancies appear especially in the bottom right quadrant.
 \black The governing equations are developed by assuming small deviations of the vortex sheet from its initial shape. In addition, a severely truncated series expansion is used to solve governing equations. Therefore, the model cannot fully capture the vortex roll-up and transition of the vortex sheet to a counter-rotating vortex pair at large times, and some discrepancies are noticeable. For non-zero values of rotation rate, an additional level of vorticity $\gamma_r$ sheds from the rotor circumference due to blade rotation. Given the sinusoidal nature of vorticity due to yaw offset \eqref{Eq:Gamma_shed_surface}, the additional shedding vorticity due to rotation increases the vorticity magnitude on either bottom or top halves of the wake (e.g., bottom half for the data shown in figure \ref{fig:uniform-that}), which in turn accelerates the vortex roll-up. At larger values of rotation rate, discrepancy is thus expected to be higher due to the earlier occurrence of vortex roll up. This is confirmed in figure \ref{fig:uniform-that} by comparing model predictions at the same dimensionless time (e.g., $\hat{t}=-1.6$), but different values of rotation rate. \black

\begin{figure}
\centerline{\includegraphics[width=.9\textwidth]{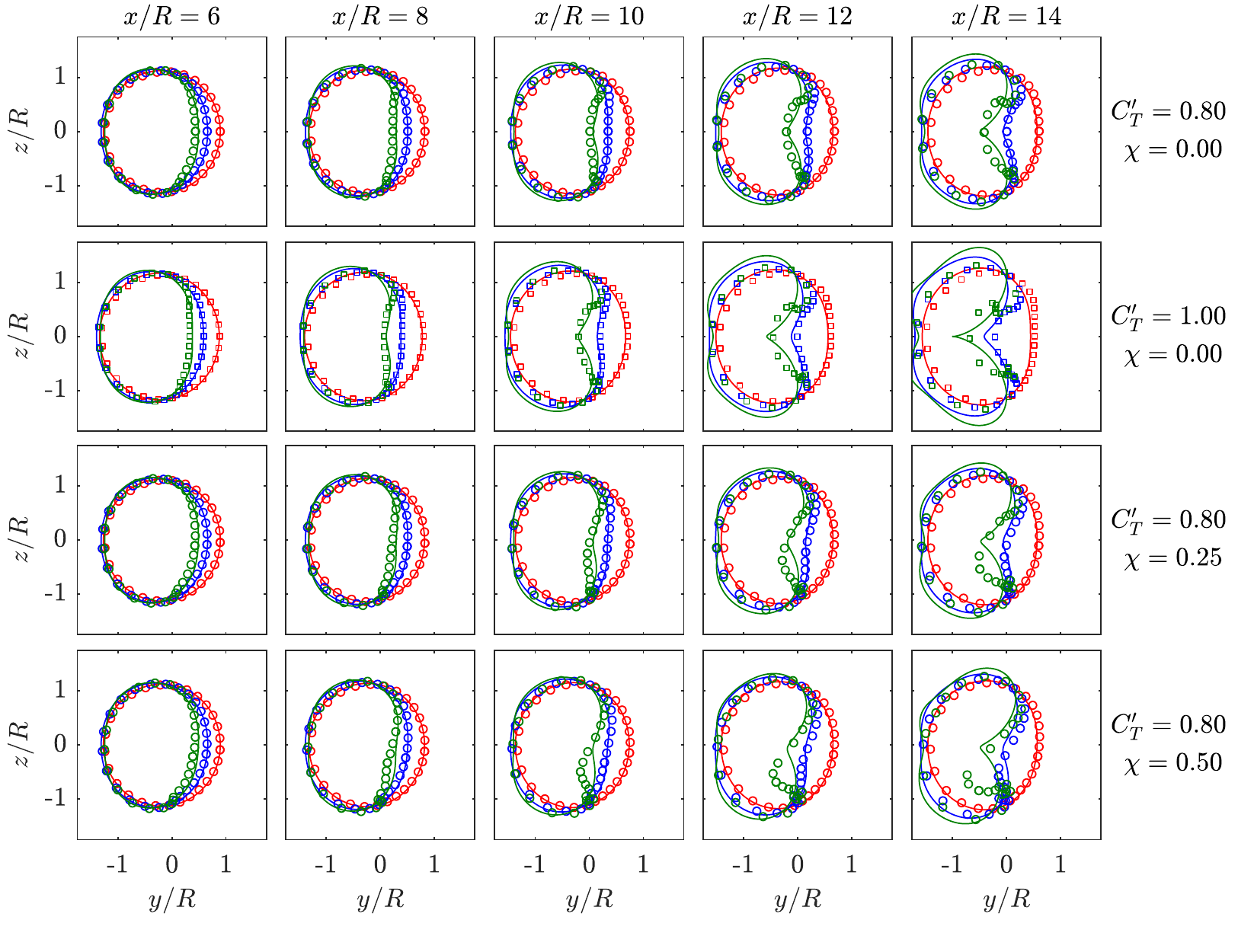}}
\caption{Wake of yawed wind turbines in uniform inflow for yaw angles $\beta=10^\circ$ (red),  $\beta=20^\circ$ (blue), $\beta=30^\circ$ (green) at various downstream locations $x/R$ and rotation rates $\upchi$. LES measurements are shown with symbols and modeled wake locations are shown with solid lines.}
\label{fig:uniform-xr}
\end{figure}

Next, we expand the comparison and plot the results in terms of physical parameters that are more directly related to the flow configuration: downstream distance $x/R$, local thrust coefficient $C_T^\prime$, yaw angle $\beta$, and dimensionless rotation rate $\upchi$. Figure~\ref{fig:uniform-xr} shows wake edge predictions of the analytical model together with the LES data for different values of $C_T'$, $\beta$ and $\upchi$, at several downwind locations. The figure shows that the degree of wake curling increases with yaw angle, thrust coefficient and streamwise distance, as expected from the analysis presented in \S\ref{sec:model_derivation}. Moreover, wake rotation  breaks the vertical symmetry of the wake. The results presented in figure~\ref{fig:uniform-xr} show that the wake shape depends strongly on all of the varied parameters: thrust coefficient, yaw angle, and rotation rate.
The analytical model developed is seen to agree well with the LES results, up to intermediate levels of wake curling.
The analytical model successfully predicts the shape of the wake for various operating conditions.  As the wake deformation grows further downstream or at increasing $C_T^\prime$, differences appear, as mentioned before due to the limitations of the model that is based on a severely truncated series expansion.  
It is clear that there is reduced agreement in the lower half of the wake for cases with large values of yaw angle and rotation rate. \black As discussed earlier, \black for these cases the lower half of the wake cross-section is subject to a strong vortex roll-up
caused by the cumulative vorticity due to both yaw offset and rotation. 
Still, the model is   able to predict many qualitative features of the wake shape, including its vertical asymmetry for cases with rotation. In addition, the sideways displacement of the entire wake is also captured quite well in all cases.

\section{Vortex sheet evolution in turbulent atmopsheric boundary layer}
\label{sec:turb_model}

In this section we generalize the prior analytical model of a yawed wind turbine wake that is applicable for ideal non-turbulent flow to the case of a  wake with a turbulent atmospheric boundary layer background flow. Here we seek a model for the entire mean velocity distribution as a function of downstream distance as well as cross-stream position, accounting for the fact that the wake will be curling due to turbine yaw.

\subsection{Vortex sheet evolution in turbulent atmospheric boundary layer}
For the ideal flow case, we assumed that the vortex sheet does not decay as it moves downstream, and the strength of the streamwise vorticity only evolves along the vortex sheet in time following idealised vortex dynamics. 
Although this may be an acceptable assumption for turbines with uniform non-turbulent inflows, it is not expected to be valid for turbines immersed in turbulent environments such as the atmospheric boundary layer (ABL). The vertically-varying mean inflow velocity in the ABL can be approximated as $U_{\rm{in}}(z) = (u_*/\kappa)\ln(z/z_0)$, where $u_*$ is the friction velocity, $\kappa$ is the von-K\`{a}rm\`{a}n constant, and $z_0$ is the roughness height. As discussed in \cite{shapiro2020decay}, the vorticity shedding from a yawed rotor decays due to the turbulent diffusion; i.e., $\gamma_b(t)$ and $\gamma_r(t)$ are functions of time. Instead of solving the full governing equations for diffusing  vortex sheet, we
approximate the effects of diffusion by
scaling $\gamma$ and the velocities by the circulation $\gamma_b(t)$ that is decaying according to a previously obtained analytical solution for the decay of CVP in a turbulent boundary layer \citep{shapiro2020decay}. Specifically, we define new scaled variables $\hat{\gamma}$,
$\hat{u}_r$ and
$\hat{u}_\theta$ such that 
\begin{align}\label{eq:dimonsionless_def_decay_gamma}
 \gamma(\theta,t) & = \gamma_b(t) \, \hat{\gamma}(\theta,\hat{t}),\;\;{u}_r(\theta,t)= \gamma_b(t)\, \hat{u}_r(\theta,\hat{t}),\;\;{u}_{\theta}(\theta,t)=\gamma_b(t) \, \hat{u}_{\theta}(\theta,\hat{t}).
 \end{align}
 If one defines scaled displacement and time  according to
 \begin{align}
\hat{\xi}(\theta,\hat{t})&=\frac{\xi(\theta,t)}{\xi_0},\\
\hat{t}(t)&=\frac{1}{\xi_0}\int\limits_0^t\gamma_b(t')\textrm{d}t',\label{eq:dimonsionless_def_decay_t}
\end{align}
one recovers the original governing equations (\ref{Eq: u_r_dimensionless}), (\ref{Eq:u_c_dimensionless}) and (\ref{Eq:empirical}) for the new definition of variables $\hat{\gamma}, \hat{u}_r, \hat{u}_\theta, \hat{\xi}$ and $\hat{t}$. An additional term is however introduced in (\ref{Eq:transport_dimensionless})
\black, so
\begin{equation}\label{eq:vort_transport_v2}
    \frac{\partial \hat{\gamma}}{\partial \hat{t}}+\frac{1}{\hat{\xi}} \frac{\partial (\hat{\gamma} \hat{u}_\theta)}{\partial \theta}\approx-   \frac{\xi_0\hat{\gamma}}{\gamma^2_b}\frac{\textrm{d} \gamma_b}{\textrm{d} t}.
\end{equation}
It will be shown later in \eqref{Eq:Gamma_b_num_integ} that $\int(\gamma_b/\gamma_{b0})\textrm{d} t \propto c(1-\textrm{exp}(k_vt/c))$, where $\gamma_{b0}=\gamma_b(t=0)$, $c$ is a constant and $k_v$ is the expansion rate of the turbulent diffusive scale, and it is modelled as $u_*/U_{in}(z)$ \citep{shapiro2020decay}. So the Taylor expansion of $\gamma_b(t)$  is given by
\begin{equation}
    \gamma_{b}(t)\approx \gamma_{b0}(1+\mathcal{O}(k_v)t+\mathcal{O}(k^2_v)t^2).
\end{equation}
For atmospheric flows, the value of $k_v$ at the hub height is equal to $u_*/U_h<\!<1$ \citep{shapiro2020decay}. Therefore, $\gamma_b$ is a slow varying reference quantity, and the additional term on the right-hand side of \eqref{eq:vort_transport_v2} is neglected for simplicity.   
\black
 This means that the solution already developed in \S\ref{sec:solve eqs- power series} will still be used as a model for the 
decaying vortex sheet once rescaled by the prescribed $\gamma_b(t)$ evolution and using the modified time $\hat{t}$.  Note that for a constant $\gamma_b$, (\ref{eq:dimonsionless_def_decay_gamma})-(\ref{eq:dimonsionless_def_decay_t}) become the same as dimensionless variables defined earlier for a non-decaying vortex sheet. 

Next, in order to evaluate the integral of (\ref{eq:dimonsionless_def_decay_t}) and derive a relationship for $\hat{t}$ under turbulent inflow conditions
we need to specify a convection velocity under turbulent inflow conditions. Due to  atmospheric turbulence, the wake mixes and recovers more quickly than in the laminar inflow case, and thus the mean velocity at the wake edge is comparable to the incoming velocity. We therefore assume that in this case the vortex sheet at a height $z$ is convected downstream with the incoming velocity at that height $U_{\rm{in}}(z) $; thus $U_{con}=U_{\rm{in}}(z)$ and $t \approx x/U_{\rm{in}}(z) $. 


To evaluate (\ref{eq:dimonsionless_def_decay_t}), we must specify the decay of vorticity, i.e. of $\gamma_b$ with streamwise distance. In \cite{shapiro2020decay} the decay of the total vortex circulation $\Gamma_b(x)$ was studied, its relationship with the density $\gamma_b$ being $\gamma_b=\Gamma_b/2R$. The resulting derived model for the total vortex circulation $\Gamma_b(x)$ as function of downstream distance $x$ is given by
\begin{equation}\label{eq:eta}
\frac{\Gamma_b(x)}{\Gamma_{b0}}=\frac{\sqrt{\pi}}{4}\frac{R}{\eta(x)}\exp\left(-\frac{R^2}{8\eta^2(x)}\right)\left[I_0\left(\frac{R^2}{8\eta^2(x)}\right)+I_1\left(\frac{R^2}{8\eta^2(x)}\right)\right],
\end{equation}
where $\Gamma_{b0}=\Gamma_{b}(x=0)$, $I_n$ is the modified Bessel function of the first kind with order $n$, and $\eta(x)=k_{\nu}(x-x_0)/24^{1/4}$ is the turbulent diffusive scale, 
and $x_0$ is the virtual origin assumed to be zero in the current work for simplicity. 

One can use $\gamma_b=\Gamma_b/2R$, $t\approx x/U_{\rm{in}}(z)$, and $\eta(x)\approx k_{\nu}x/24^{1/4}$ to rewrite (\ref{eq:dimonsionless_def_decay_t}) as
\begin{equation}\label{Eq:t_hat_turb_v2}
    \hat{t}=\frac{24^{1/4}}{2k_{\nu}U_{\rm{in}}(z)\xi_0 R }\int\limits_0^\eta\Gamma_b(\eta')\textrm{d}\eta'.
\end{equation}
Numerical integration of $\Gamma_b$, expressed by (\ref{eq:eta}), yields results that can be conveniently approximated by the following (fitted) expression:  
\black
\begin{equation}\label{Eq:Gamma_b_num_integ}
    \frac{1}{\Gamma_{b0} R}\int\limits_0^\eta\Gamma_b(\eta')\textrm{d}\eta'\approx 1.3\left[1-\textrm{exp}\left(-\frac{\eta(x)}{1.3R}\right)\right].
\end{equation}
\black
We then use (\ref{Eq:gamma_b_final}) to express $\Gamma_{b0}$ as a function of operating conditions, approximate $\xi_0(\theta)$ with $\tilde{\xi}_0$ given by ($\ref{Eq:xi_0-circular}$), and insert (\ref{Eq:Gamma_b_num_integ}) into (\ref{Eq:t_hat_turb_v2}) to obtain
\black
\begin{equation}
\label{Eq:t_hat_decay_v3}
   \hat{t}(x,z)\approx -1.44 \, \frac{U_h}{u_*}\frac{R}{\tilde{\xi}_0} \,C_T \cos^2\beta\sin\beta \,\left[1-\textrm{exp}\left(-0.35\frac{u_*}{U_{\rm{in}}(z)}\frac{x}{R}\right)\right].
\end{equation}
\black
Due to vorticity decay, $\hat{t}$ for turbulent inflow cases increases at a slower rate than the one for laminar inflow cases. For instance, according to \eqref{Eq:t_hat_decay_v3}, for a turbine with $C_T'=1.33$ subject to an ABL with $k_\nu = 0.05$, the streamwise position associated with $|\hat{t}|=2$ varies between $17R$ and $61R$ for $\beta=30^{\circ}-10^{\circ}$. As mentioned in \S\ref{sec:results_laminar}, for the same turbine with $\beta=30^{\circ}-10^{\circ}$ subject to a laminar flow, $|\hat{t}|=2$ at $x=12R-29R$. \black It is also worth noting that the above definition of $\hat{t}$ \eqref{Eq:t_hat_decay_v3} is reduced to the one used for non-decaying vortex sheets (i.e., $\hat{t}=\gamma_{b0}t/\tilde{\xi}_0$) as $u_*$ tends to zero\black.

\begin{figure}
\centerline{\includegraphics[width=.5\textwidth]{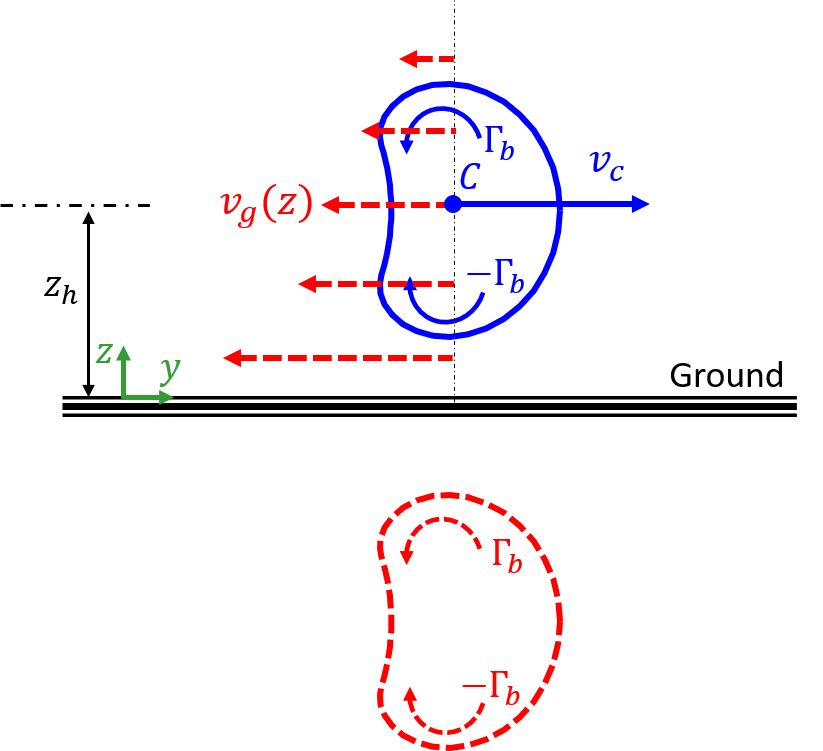}}
\caption{\black Schematic of modelling the effect of ground using an image technique.}
\label{fig:ground}
\end{figure}

The effect of the ground on the wake deflection was not modelled in the uniform inflow cases. To model the effect of the ground, we use an image technique to modify the wake centre location $y_c$, \black as shown in figure \ref{fig:ground}\black. Modelling the vortex sheet as a CVP, the image CVP induces a lateral velocity in the opposite direction, termed as $v_g$ given by
\begin{equation}
    v_g=\frac{\Gamma_b}{2\pi} \left[\frac{1}{z+z_h-\xi_0} -\frac{1}{z+z_h+\xi_0}\right] = \frac{\Gamma_b \xi_0}{\pi\left[(z+z_h)^2-\xi_0^2\right]}.
\end{equation}
Therefore, the lateral wake deflection caused by the ground is given by
\begin{equation}
    y_g(z) = \int_0^t v_g(z,t') \, dt' = \frac{\xi_0 \int_0^t \Gamma_b(t') \, dt}{\pi\left[(z+z_h)^2-\xi_0^2\right]}.
\end{equation}
Approximating $\Gamma_b \approx 2\xi_0\gamma_b$ and using $\hat{t} = \int_0^t \gamma_b(t') / \xi_0 \, dt'$ we find
\begin{equation}\label{eq:y_g_2}
    \hat{y}_g = \frac{2}{\pi} \frac{\hat{t}}{\left[  ({z+z_h)}/{\xi_0}\right]^2-1},
\end{equation}
where $\hat{y}_g=y_g/\xi_0$. \black Substituting $\xi_0$ with $\tilde{\xi}_0$ in \eqref{eq:y_g_2} for simplicity and \black subtracting this result from~\eqref{Eq:empirical} yields
\begin{equation}\label{eq:y_c_with_ground}
    \hat{y}_c = \frac{(\pi-1)|\hat{t}|^3+2\sqrt{3}\pi^2\hat{t}^2+48(\pi-1)^2|\hat{t}|}{2\pi(\pi-1)\hat{t}^2+4\sqrt{3}\pi^2|\hat{t}|+96(\pi-1)^2}\textrm{sgn}(\hat{t}) - \frac{2}{\pi}\frac{\hat{t}}{\left[({z+z_h})/{\tilde{\xi}_0}\right]^2-1}.
\end{equation}
It is worth mentioning that for $z_h\!\to\!\infty$, the second term on the right-hand side of (\ref{eq:y_c_with_ground}) vanishes, and thus the equation is reduced to \eqref{Eq:empirical}.

\subsection{Analytical model for mean velocity distribution}

The shape of the wake edge predicted earlier can now be used to model the spatial distribution of the velocity  deficit in the curled wake at each streamwise position.  For non-curled (i.e., non-yawed) wakes a number of wake profiles have already been proposed in the literature, including top-hat~\citep{Katic1986, Frandsen2006}, Gaussian~\citep{bastankhah2014new, bastankhah2016experimental}, double Gaussian~\citep{schreiber2020brief}, and super-Gaussian~\citep{shapiro2019paradigm, blondel2020}. Most of these conserve flux of momentum deficit only in its linearized version valid far downstream, see discussion in \cite{bastankhah2014new}. Here we demonstrate the use of the shape deformation for an analytical model in the context of the Gaussian wake model~\citep{bastankhah2014new, bastankhah2016experimental}, but it could be implemented in other wake models as well.

The modeled streamwise mean velocity
\begin{equation}
    U(\mathbf{x}) = U_{\rm{in}}(z) - \Delta U(\mathbf{x}),
\end{equation}
is defined based on the incoming velocity field $U_{\rm{in}}(z)$ and the modeled velocity deficit in the wake, $\Delta U(\mathbf{x})$. In the Gaussian model, the velocity deficit profile is modeled as  
\begin{equation}\label{eq:Gauss_velocity_profile}
    \frac{\Delta U}{U_h} = C(x) \exp \left[ - \frac{(y-y_c)^2 + (z-z_h)^2}{2\sigma^2} \right],
\end{equation}
where $C(x)$ is the normalised maximum velocity deficit at each streamwise location and $\sigma$ is the characteristic wake width. In (\ref{eq:Gauss_velocity_profile}), the wake centre location is $y_c=\hat{y}_c\xi_0$, where $\hat{y}_c$ is obtained from (\ref{eq:y_c_with_ground}) and $\xi_0$ is approximated with $\tilde{\xi}_0$ given in (\ref{Eq:xi_0-circular}).  In prior versions of the Gaussian wake model, it is assumed that the  characteristic wake width depends only on downstream distance. Moreover, it is assumed that it grows linearly downstream at a rate $k$. The linear growth of the wake arises from the similarity solution when eddy-viscosity is assumed to scale with a constant velocity, friction velocity $u^*$, and the wake scale $\sigma$ itself \citep{shapiro2019paradigm}, i.e.
\begin{equation}\label{Eq:sigma}
    \sigma(x) = kx+\sigma_0,
\end{equation}
where $k$ is the wake expansion rate and $\sigma_0$ the initial wake size. We now propose to include wake curling and deformation by making $\sigma$ dependent also on the angle $\theta$ according to
\begin{equation}\label{Eq:sigma}
    \sigma(x,\theta) = kx+0.4 \,\xi(\theta,x),
\end{equation}
where
 $\xi(x,\theta)=\xi_0(\theta)\hat{\xi}(\theta,\hat{t})$, and $\xi_0(\theta)$ is given by ($\ref{Eq:xi_0}$), and the dimensionless wake shape $\hat{\xi}(\theta,\hat{t})$ is given by either the analytical relation of (\ref{eq:final_xi}) for $|\hat{t}|\leq2$ or the empirical relation \eqref{eq:xi_empirical} for given values of polar angle $\theta$ and dimensionless time $\hat{t}$. The polar angle is determined at each position from $\tan\theta=(z-z_h)/(y-y_c)$, and $\hat{t}$ is given by (\ref{Eq:t_hat_decay_v3}). In (\ref{Eq:sigma}), the first term on the right-hand side of the equation expands the wake in all radial directions due to turbulent mixing, while the second one deforms the wake cross-section according to the vortex sheet solution derived earlier. According to (\ref{Eq:sigma}), for an unyawed turbine, the initial characteristic wake width is reduced to $0.4R\sqrt{A_*}$, which is the same as the one suggested by \citet{bastankhah2014new}. For the wake expansion rate, we assume $k = \alpha u_*/U_{\rm{in}}(z)$ \citep{shapiro2019paradigm}, where $\alpha$ is an empirical constant. Alternatively, $k$ can be estimated based on the turbulence intensity $I$ of the incoming boundary layer flow (i.e., $k=\alpha' I$, where $\alpha'$ is an empirical constant)  as suggested in prior studies \citep[see][among others]{Niayifar2016,carbajo2018wind,Zhan_Iungo2020}. At each height $z$, turbulence intensity $I(z)$ is defined as $\sqrt{\overline{u^{'2}}}/U_{in}(z)$, where $u'$ is turbulent fluctuation of streamwise velocity, and overbar denotes time averaging. Note that invoking the logarithmic law for the fluctuating velocity variance in high Reynolds number turbulent boundary layers \citep{marusic2013logarithmic,meneveau2013log_law}, one can show that $u_*/U_{in}$ and $I$ are related to each other by $I=(u_*/U_{in}) \left[B_1-A_1\ln(z/\delta)\right]^{1/2}$, where $\delta$ is the boundary-layer thickness, and $A_1$ and $B_1$ are constants.  



The maximum velocity deficit $C(x)$ in (\ref{eq:Gauss_velocity_profile}) is obtained by enforcing the conservation of streamwise momentum deficit flux $\rho \int \Delta U (U_h - \Delta U) \, dA \approx T \cos \beta$. To simplify the integration and avoid dependence on $\theta$, we approximate $\sigma^2(x,\theta)$ with $\tilde{\sigma}^2(x)$ where the latter is given by 
\begin{equation}\label{eq:sigma_tilde}
    \tilde{\sigma}^2(x) = (kx + 0.4\tilde{\xi}_0)(kx + 0.4\tilde{\xi}_0\cos\beta).
\end{equation} 
In this expression, $\tilde{\xi}_0$ is given by (\ref{Eq:xi_0-circular}) and $k=\alpha u_*/U_h$ is the wake expansion rate at $z=z_h$. This yields
\begin{equation}\label{eq:max_deficit}
    C(x) = 1 - \sqrt{1 - \frac{C_T \cos^3 \beta}{2 \tilde{\sigma}^2(x)/R^2}}.
\end{equation}
Note that in stating conservation of flux of streamwise momentum deficit, pressure and turbulent and viscous shear stress effects are assumed to be negligible. This may be a questionable assumption in the near wake region as well as far wake of turbines deep inside a wind farm \citep{bastankhah2020cumulative_wake}.

For the sake of completeness, a summary of the steps required to implement the proposed model and predict wake velocity deficit distributions at a given downwind location $\textbf{x}=(x,y,z)$ is provided below:
  (1) Compute the approximate form of the initial wake shape $\tilde{\xi}_0$ (\ref{Eq:xi_0-circular}).
  (2) Determine the dimensionless time $\hat{t}$ from (\ref{Eq:t_hat_decay_v3}).
  (3) Find the wake centre location $y_c\approx\hat{y}_c\tilde{\xi}_0$, where $\hat{y}_c$ is given by (\ref{eq:y_c_with_ground}).
  (4) Find the polar angle $\theta$, which is measured from the positive $y$-axis toward the positive $z$-axis such that $\tan\theta=(z-z_h)/(y-y_c)$.
  (5) Evaluate the initial wake shape $\xi_0(\theta)$ (\ref{Eq:xi_0}).
    (6) Calculate the wake shape function $\xi(\theta,x)=\xi_0(\theta)\hat{\xi}(\theta,\hat{t})$, where $\hat{\xi}(\theta,\hat{t})$ can be estimated either from the analytical solution (\ref{eq:final_xi}) (if $|\hat{t}|<2$) or the empirical one \eqref{eq:xi_empirical}, and $\upchi$ is given by (\ref{eq:rotation_rate}).
  (7) Find the wake width $\sigma(x,\theta)$ based on (\ref{Eq:sigma}).
  (8) Evaluate the maximum velocity deficit $C(x)$ from (\ref{eq:max_deficit}), where $\tilde{\sigma}^2(x)$ is given by (\ref{eq:sigma_tilde}).
  (9) Determine the wake velocity deficit $\Delta U$ according to (\ref{eq:Gauss_velocity_profile}).

\subsection{Comparison with large eddy simulations} \label{sec:turb_result}

In the following, the  streamwise mean velocity distribution based on the proposed model is compared with the LES data for turbulent ABL inflow cases. Simulations of yawed wind turbines represented as \black rotating \black actuator disks \black (ADM-R) \black are performed using  $C_T' = 1.33$ \black and a local tip speed ratio of $\lambda'=10.67$ \black at yaw angles of $\beta = 15^\circ$, $20^\circ$, $25^\circ$, and $30^\circ$~\citep{shapiro2020decay}, \black where the local tip speed ratio is defined as $\lambda'=\Omega R/u_d$ \black. \black In unyawed conditions, the selection of local thrust coefficient and tip speed ratio corresponds to  $C_T = 0.75$ and $\lambda=8$, which are realistic of modern utility-scale turbines. \black The actuator disk with diameter $D = 100$ m and a hub height of $z_h = 100$ m is placed 500 m from the inlet of a domain with an effective size of $L_x = 3.75$ km, $L_y = 3$ km, and $L_z = 1$ km. 
The domain is divided into $N_x = 360$, $N_y = 288$, and $N_z = 432$ grid points. The velocity field is averaged for a time $\mathcal{T} u_*/L_z \approx 8$, where $u_* = 0.45$ m/s. 
A $0.49L_z$ shift is used to reduce streamwise streaks in the time-averaged velocity field. The roughness height is $z_0 = 0.1$ m. 

\begin{figure}
\centerline{\includegraphics[width=\textwidth]{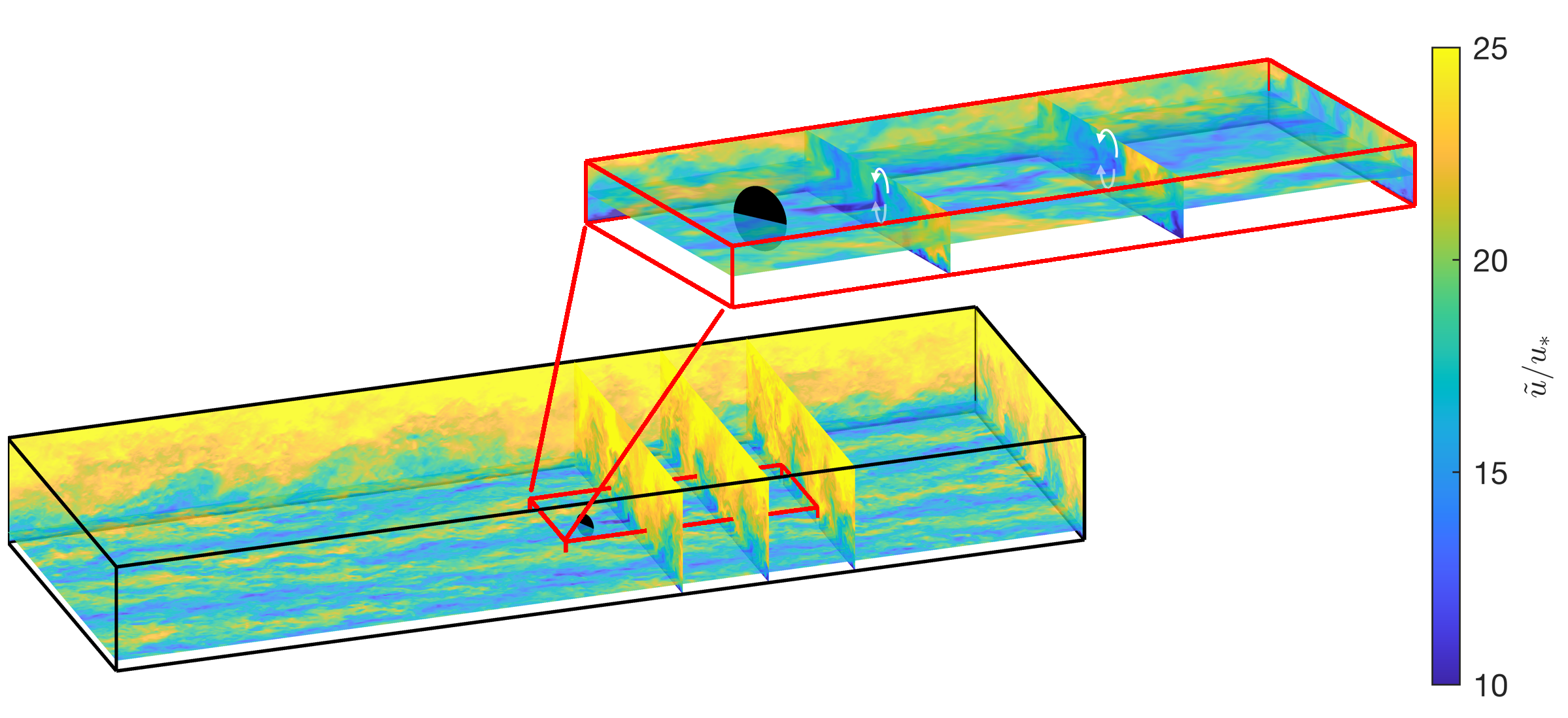}}
\caption{Contour plots of instantaneous streamwise velocity including a wind turbine with turbulent boundary layer inflow from LES. Turbine operating parameters are $C_T' = 1.33$ and yaw angle $\beta=25^\circ$. Contours are shown through the turbine centre at $z=z_h$, at the back of the domain at $y=L_y$ and $x=L_x$ and at cross planes of $x/R=8$, $24$, and $40$. The swept area of the rotor is denoted as a black circle. A zoomed in flow field around the turbine (red box) is also shown and white arrows highlight the sense of rotation of the induced counter rotating vortex pair.}
\label{fig:snapshot}
\end{figure}

Figures~\ref{fig:snapshot} show contour maps of streamwise velocity on representative planes across the LES domain. It shows the streaks in the turbulent atmospheric boundary layer at the turbine hub height and the generated wake behind the yawed wind turbine, which is shown as a black circle. At downstream locations, the effect of the the curled yawed wind turbine wake on the cross-plane velocity field is shown at downstream locations of $x/R = 8$, $24$, and $40$.

\begin{figure}
\centerline{\includegraphics[width=\textwidth]{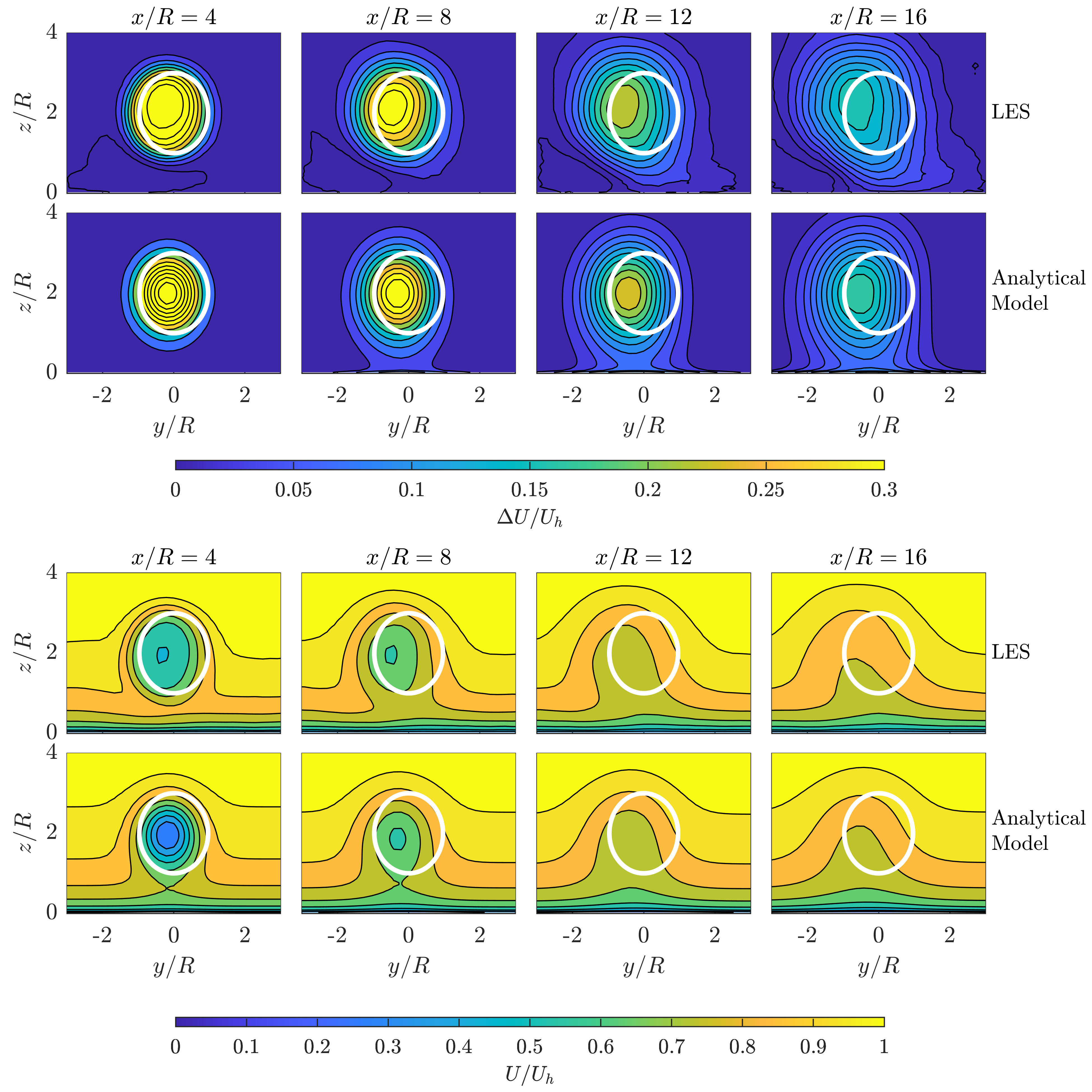}}
\caption{Top figures: Contour plots of normalised wake velocity deficit behind a wind turbine in turbulent inflow with a thrust coefficient of $C_T' = 1.33$ \black and local tip speed ratio $\lambda'=10.67$ \black at a yaw angle of $\beta=15^\circ$. White circles indicate frontal area of wind turbines. Bottom figures:  Contour plots of normalised streamwise velocity behind the same turbine.}
\label{fig:turbulent-rot-xr-contour-beta15}
\end{figure}

\begin{figure}
\centerline{\includegraphics[width=\textwidth]{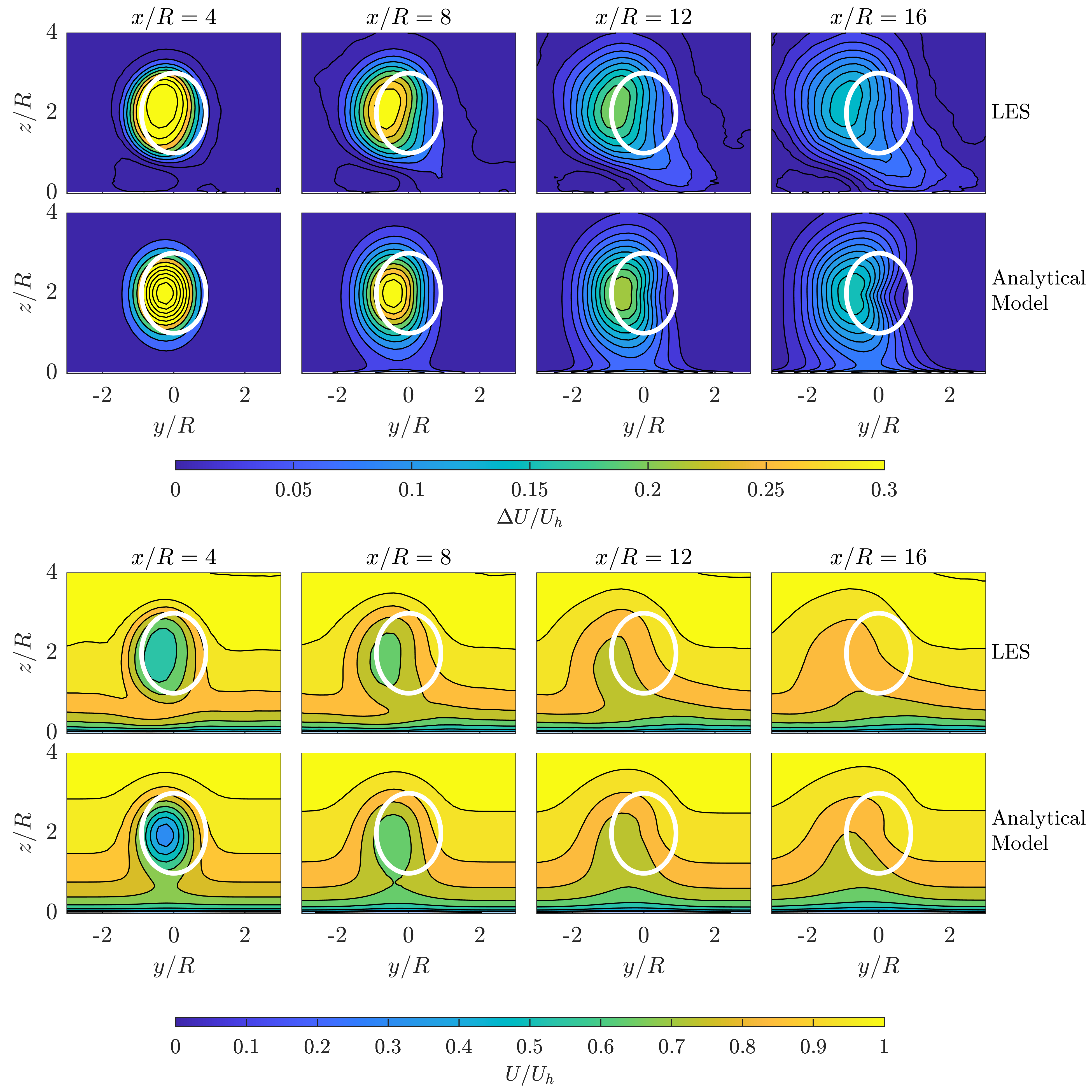}}
\caption{Same as figure \ref{fig:turbulent-rot-xr-contour-beta15} but for a yaw angle of $\beta=25^\circ$.}
\label{fig:turbulent-rot-xr-contour-beta25}
\end{figure}

Figures~\ref{fig:turbulent-rot-xr-contour-beta15} and \ref{fig:turbulent-rot-xr-contour-beta25} show the wake mean velocity distributions based on the LES results and the analytical model for 
$\beta = 15^\circ$ and $25^\circ$ at various downstream locations. Top panels show the normalised velocity deficit $\Delta U/U_h$ and bottom panels show the normalised streamwise velocity distribution. 

As in past studies~\citep{bastankhah2016experimental,shapiro2018}, the wake recovery rate $k$ in the analytical model is calculated by fitting a Gaussian profile to the downstream wake profiles at $z=z_h$. This gives the resulting wake expansion rate as $k=0.6\, u_*/U_{in}$. As seen in figures~\ref{fig:turbulent-rot-xr-contour-beta15} and \ref{fig:turbulent-rot-xr-contour-beta25}, the model captures the curling and deflection of the wake as well as some variation in the wake deflection as a function of vertical distance due to ground effects. The LES results display \black slightly \black less curling than the model as well as more noticeable wake deflection towards the ground that is opposite to the deflection of the bulk of the wake. \black The analytical model does not also predict small vertical wake deflections observed in the LES data. According to the vortex sheet analysis performed in the current study (see \eqref{Eq:z_c_eq2}), this vertical deflection is due to rotation effects (i.e., non-zero values of rotation rate $\upchi$).  As mentioned in \S\ref{sec:lateral_deflection}, for simplicity we did not include the vertical wake deflection in the final version of the model. 
Despite these small differences, figures~\ref{fig:turbulent-rot-xr-contour-beta15} and \ref{fig:turbulent-rot-xr-contour-beta25} show that overall model predictions are in acceptable agreement with the LES data. \black

Wake flow results are also used to compute the power of a downwind turbine based on both the  proposed analytical model and the LES data. In both cases, a virtual wind turbine is placed in the flow field at different distances downstream of the yawed turbine. The idea is that as yawing increases, the downstream turbine will overlap less and less with the wake due both to the sideways displacement of the wake and to the curled crescent shaped wake that creates lower velocity deficit at the centre of the hypothetical downstream  turbine. 

In order to evaluate the power generated by the downwind turbine, we require the disk-averaged streamwise velocity defined according to
\begin{equation}
    U_d = (1-a)\frac{1}{\pi R^2} \iint \limits_{\rm disk} U(x_T,y,z) \, dz dy,
\end{equation}
where  
$U(x_T,y,z)$ is the mean velocity in the flow at the turbine location $x_T$ computed from the model or from the LES and the integration covers the turbine disk area. For the latter case, we evaluate the mean velocity by time averaging: $U(x_T,y,z)=\langle \tilde{u}_1(x_T,y,z)\rangle$. Since the turbine is not included in the simulation, the turbine disk velocity that would occur there includes the $(1-a)$ prefactor, where  $a$ is the turbine's assumed induction factor. The power is subsequently calculated as
\begin{equation}
    P = \frac{1}{2} \rho \pi R^2 C_T' U_d^3 ,
\end{equation}
and it is normalized by the power that an unyawed free-standing turbine would generate under similar conditions, $P_0 = (1/2) \rho \pi R^2 C_T'(1-a)^3 U_h^3$ (thus making the result independent of assumed $C_T'$, etc.). 

\begin{figure}
 \begin{minipage}[c]{0.40\linewidth}
 \centerline{
  \includegraphics[width=\linewidth]{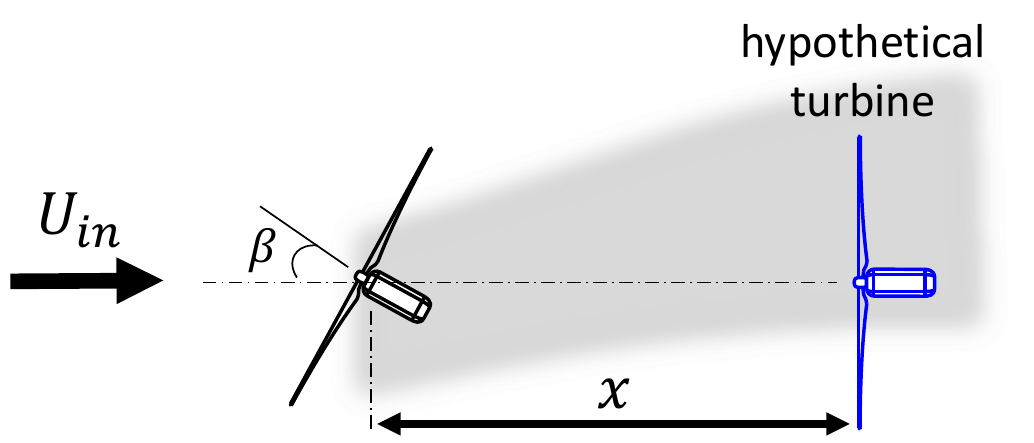}}
  \vspace{3em}
 \end{minipage}
 \hfill
 \begin{minipage}[c]{0.55\linewidth}
  \includegraphics[width=.75\linewidth]{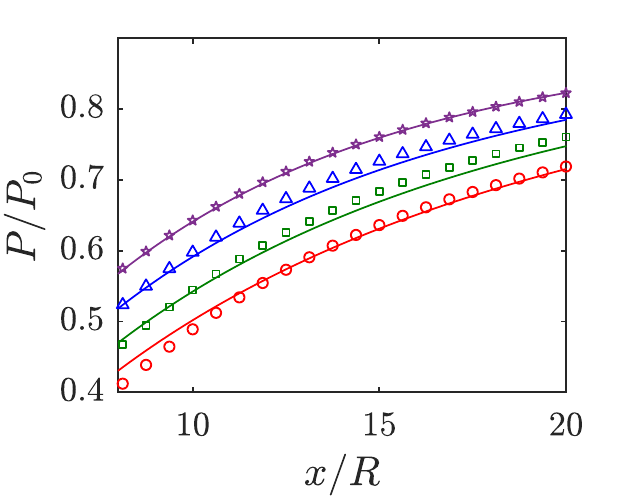}
 \end{minipage}%
 \caption{Left: sketch of a hypothetical turbine placed at various locations downstream of the yawed turbine in the LES field. Right: Normalized power of the hypothetical turbine operating at $C_T' = 1.33$. Different yaw angles in LES are $\beta=15^\circ$ ({\color{red}$\circ$}),  $\beta=20^\circ$ ({\color{matlab-green}$\square$}),  $\beta=25^\circ$ ({\color{blue}$\triangle$}), and $\beta=30^\circ$ ({\color{matlab-purple}\faStarO}). The predictions based on the analytical curled wake model are shown as solid lines.}\label{fig:power}
\end{figure}

Figure~\ref{fig:power} shows the normalised power of this hypothetical turbine as a function of streamwise spacing for different yaw angles of the upwind turbine.  The figure demonstrates good agreement between the analytical model and the LES results for a broad range of streamwise spacings and yaw angles.

\subsection{Comparison with experimental data} \label{sec:comp_exp}
\begin{figure}
\centerline{\includegraphics[width=1\textwidth]{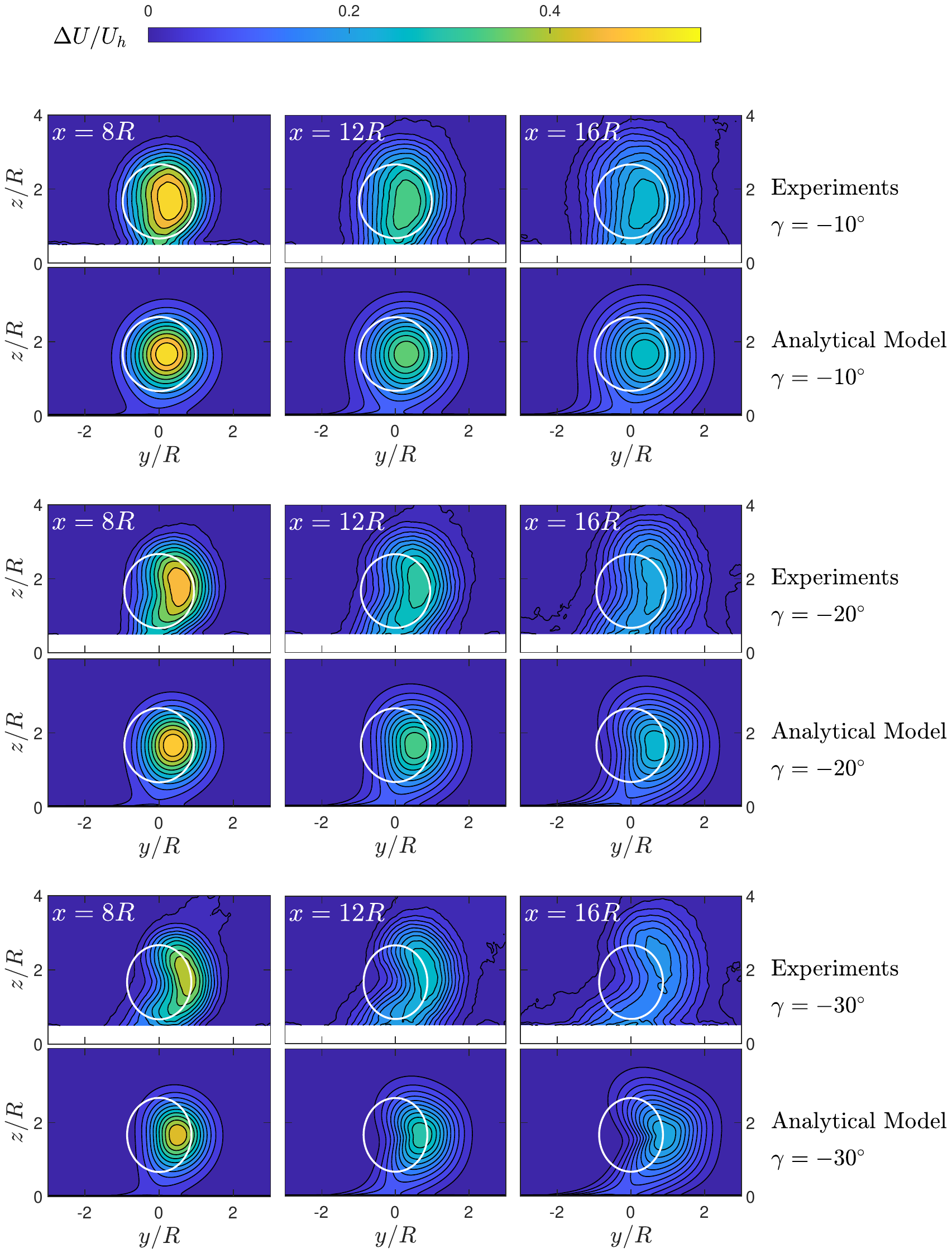}}
\caption{Contours of normalised velocity deficit in $yz$ planes at different downwind locations and different yaw angles based on: wind-tunnel experiments \citep{bastankhah2016experimental}, and the new proposed analytical model. White circles indicate the frontal area of wind turbine.}
\label{fig:exp_analytic}
\end{figure}

Model predictions are also compared with wind-tunnel experiments by \citet{bastankhah2016experimental}. Flow measurements were performed to quantify the wake of a yawed wind turbine with a diameter of $15$ cm and hub height of $12.5$ cm, and the turbine is subject to a turbulent boundary layer, naturally developed over the smooth surface of the wind-tunnel floor. Additional information on turbine properties ($C_T$, $\lambda$, etc.) and inflow conditions ($U_h$, $I$, etc.) may be found in 
\citet{bastankhah2016experimental}. The wake recovery rate $k$ for the analytical model is estimated based on $k=0.35I$ \citep{carbajo2018wind}. Figure \ref{fig:exp_analytic} shows contours of normalised velocity deficit in $yz$ planes at different downwind locations and different yaw angles based on both experiments and model predictions. Overall the figure shows that the proposed model is able to successfully predict the complex curled shape of the wake and its lateral deflection. The opposite wake deflection close to the ground is also well captured by the model.  While the overall trends and qualitative features of the velocity defect distribution of the curled wake are reproduced by the analytical model, some differences between measurements and model can still be discerned. The figure shows that the vertical extent of the lower half of the wake is underestimated by the analytical model, which is mainly due to neglecting the wake of the turbine tower in the analytical model. Tower wake effects are however expected to be less significant for utility-scale wind turbines which tend to have less bulky towers (with respect to the rotor diameter).

\section{Summary and conclusions} \label{sec:summary}

A curled shape is a typical aerodynamic feature of wakes behind  yawed turbines.  In this study, we develop an analytical model to describe the wake shape and its downstream evolution for both uniform ideal inflow and turbulent boundary layer background flow. The predicted wake shape is then used in a wake model to describe analytically the mean velocity deficit distribution behind a yawed wind turbine.

To model the curled shape of the wake, we represent the wake edge as a vortex sheet shedding from the rotor disk circumference due to both yaw offset and rotation. 
A simple relationship is developed to estimate the initial distribution of vorticity along this vortex sheet as a function of turbine operating conditions such as thrust coefficient, yaw angle and tip-speed ratio. The goal is to obtain the evolution of the locus of this vortex sheet with time. The governing equations for the deformation of the vortex sheet are developed based on the Biot-Savart law and the vorticity transport equation. After non-dimensionalising the equations, they are solved using a power series expansion method. Subsequently, assuming a given downstream convection velocity, we  map the time evolution to a spatial one in the streamwise direction.  
The developed solution is only valid for limited range of dimensionless times ($|\hat{t}|< 2$) or equivalent distances. For larger values of $|\hat{t}|$, an empirical expression is also proposed, which complies with the derived analytical solution for smaller times but is also realistic for larger times.

Apart from deforming to a curled shape, the vortex sheet is also deflected laterally behind a yawed turbine. This deflection is modelled with an equation  combining analytically derived deflections using two approaches: considering an approximately circular vortex sheet (valid for small times) and considering self-induced motion of a CVP (valid for larger times). Moreover,  close to the ground the wake is deflected in the opposite direction. This ground effect is modelled using image flow that introduces an additional deflection term to account for the velocity induced by the image CVP. 


Wake shape predictions are first compared with numerical simulation data for a yawed turbine placed in a  uniform, non-turbulent inflow. 
Several cases with different values of thrust coefficients, yaw angles and tip-speed ratios are considered. It is shown that the analytical model predictions agree well with LES results at moderate times. Also in agreement with the theory, we show that the numerical simulation results can be collapsed into a common  wake shape when the problem variables are normalised using the scaling as suggested by the theoretical model.  

The theory is then adapted to the case of turbulent ABL inflow for applications to real wind turbine flows. It is known that, unlike the case of ideal flow,  turbulent diffusion plays a central role in weakening the streamwise vortices during their downstream evolution. 
This phenomenon is modeled by  non-dimensionalizing   the governing equations using a time-varying reference vortex sheet strength. 
Finally, we modify the Gaussian wake model 
to incorporate the predicted shape $\xi(\theta,\hat{t})$ and lateral deflection $y_c$ of the curled wake. The analytical curled wake model thus describes the full spatial distribution of mean streamwise velocity downstream of a yawed turbine.  To validate the model predictions for the ABL case,  wake velocity contours at different downstream positions are compared with  LES and wind-tunnel results at various yaw angles. Moreover, the power extracted by a hypothetical wind turbine located at different downwind positions is computed and compared to similar results from the LES-generated mean velocity distributions. A good agreement of the model predictions with the numerical and experimental data is observed, suggesting that the proposed model includes the most relevant fluid mechanical effects governing the mean velocity distribution in wakes downstream of yawed wind turbines.  Also, the proposed analytical model should be useful for tasks such as wind farm optimization and control, where numerical simulations tend to be too time consuming and costly. \black While to our knowledge the developed analytical model is the first of its kind to predict the curled shape of yawed turbine wakes, more research is still needed to shed light on the impact of ABL characteristics such as wind veer on the wake of a yawed turbine. Of special interest is to study the combined effect of yaw offset and wind veer on the wake cross-section in future works. Moreover, the effect of ground and rotation on the wake cross-section needs to be thoroughly studied in future works for turbines with different geometries and operating conditions. \black   

 
\section*{Acknowledgements}
MB acknowledges funding from Innovate UK (grant no. 89640). CS, DG and CM acknowledge funding from the National Science Foundation (grant nos. 1949778) and computational resources from MARCC and Cheyenne (doi:10.5065/D6RX99HX). 
 
\section*{Declaration of interests}
The  authors  report  no  conflict  of  interest.
\appendix
\section{Evaluation of principal value of required integrals}
\label{appA}

This appendix evaluates the integrals (\ref{eq:pv_int1}) and (\ref{eq:pv_int2}). Here, we only provide the proof for (\ref{eq:pv_int1}), as (\ref{eq:pv_int2}) can be solved similarly. First, we define $I$ as 

\begin{dmath}\label{Eq:1}
I \coloneqq \int_{0}^{2\pi} \frac{\sin(n x)}{\tan{\left[\left(x-b\right)/2\right]}} \, \textrm{d}x.
\end{dmath}
By defining the complex parameters $i \coloneqq \sqrt{-1}$ and $ a \coloneqq e ^ {-i \frac{b}{2}} $ and the complex variable $z \coloneqq e^{ix}$, and considering that $\textrm{d}x=\textrm{d}z/(iz)$, $\sin(n x)={( z^{n} - z^{-n} )}/{(2i)}$ and $\tan[{(x-b)}/{2}]= -i {( a^2 z-1 )}/{( a^2 z+1 )}$, the integral $I$ can be written as:

\begin{dmath}\label{Eq:8}
I= \frac{1}{2 i} \int_{C_1 } F(z) \, \textrm{d}z,
\end{dmath}
where $C_1$ is the integration path, which is the unit circle, $|z|=1$, on the complex plane, and $F(z)$ is defined as:

\begin{equation}\label{Eq:8-1}
F(z)\coloneqq \frac{ (z^{2n} - 1) (a^2 z+1) }{ a^2 z^{n+1}(z-\frac{1}{a^2})}.
\end{equation}
In order to evaluate the integral in (\ref{Eq:8}), we need to calculate the residues of $F(z)$ at its singularities. $F(z)$ has two singularities: one at $z=0$ (pole of order $n+1$) and the other at $z=a^{-2}$ (pole of order one). To calculate the residue of $F(z)$ at $z=0$, we define a function $\Phi(z)$ such that \citep[see][chapter 4]{Ablowitz2003}:

\begin{equation}\label{Eq:9}
F(z)= \frac{\Phi(z)}{z^{n+1}}.
\end{equation}

Therefore the residue at $z=0$ is:

\begin{dmath}\label{Eq:10}
\textrm{Res}( F(z); 0) = \left[ \frac{1}{n!} \frac{\textrm{d}^n}{\textrm{d}z^n} \Phi(z) \right]_{z=0},
\end{dmath}
or
\begin{dmath}\label{Eq:11}
\textrm{Res}( F(z); 0) = \frac{1}{n!} \left[ \sum\limits_{k=0}^n {n \choose k} \frac{\textrm{d}^{n-k}}{\textrm{d}z^{n-k}}(a^2 z^{2n+1} + z^{2n} -a^2 z  -1) \  \frac{\textrm{d}^{k}}{\textrm{d}z^{k}} \frac{1}{a^2 z - 1 } \right]_{z=0}.
\end{dmath}
It can readily be shown that the first derivative term is non-zero only for $k=n-1 $ and $k=n$. Also expanding the second derivative, and after some manipulation, we obtain:

\begin{equation}\label{Eq:15}
\textrm{Res}( F(z); 0) = 2a^{2n} = 2 e^{-i b n}.
\end{equation}

Now we calculate the residue of $F(z)$ at $z=a^{-2}$. To do this, we define a function $\Phi'(z)$ such that:

\begin{equation}\label{Eq:16}
F(z) =  \frac{\Phi'(z)}{(z-\frac{1}{a^2})}.
\end{equation}
Therefore the residue at $z=a^{-2}$ is:
\begin{equation}\label{Eq:17}
\textrm{Res}( F(z); a^{-2}) = \Phi'(z) (z=a^{-2}) = 2(e^{i b n} - e^{-i b n}) .
\end{equation}

Since the second singularity is located on the integration path, we cannot use Cauchy Residue Theorem directly. However, this integral can be regarded as a Cauchy Type Integral, and we can use the Plemelj Formulae to evaluate its Cauchy Principal Value. To do this, we assume $I^+$ is the limiting value of the integral $I$ when the second singularity of $F(z)$ (i.e., $z=a^{-2}=e^{ib}$) approaches the integration path ($|z|=1$) from inside the unit circle, and $I^-$ is the limiting value of the integral $I$ when the second singularity of $F(z)$ approaches the integration path from outside the unit circle \citep[see][chapter 7]{Ablowitz2003}. Thus:
\begin{equation}\label{Eq:18}
I^+ = (2\pi i) (\frac{1}{2i})\left[ \textrm{Res}( F(z); 0) + \textrm{Res}( F(z); a^{-2}) \right] = 2\pi e^{i b n},
\end{equation}
and
\begin{equation}\label{Eq:19}
I^- = (2\pi i) (\frac{1}{2i})\left[ \textrm{Res}( F(z), 0) \right] = 2\pi e^{-i b n}.
\end{equation}

According to the Plemelj Formulae, we have the following for the principal value of integral $I$:
\begin{equation}\label{Eq:19}
\textrm{p.v.}(I) = \frac{1}{2}(I^+ + I^-) = \frac{1}{2} (2\pi e^{i b n} + 2\pi e^{-i b n} ) = 2 \pi \cos{(b n)},
\end{equation}
and (\ref{eq:pv_int1}) is proved. 


\section{Empirical vortex sheet shape model for large times}\label{sec:appendix_empirical}
\label{sec:empirical_relation}





The analytical solution for the deformation of the vortex sheet derived earlier is 
valid only for relatively short times.
In fact, $\xi(\theta,\hat{t})$ can become  negative (the vortex sheet crosses itself and becomes non-simple and non-analytic) at dimensionless times with magnitude larger than a critical time denoted by $\hat{t}_{c}$. 
Therefore, the analytical solution should be only used for $|\hat{t}|<\hat{t}_c$. By setting $\hat{\xi}=0$ in (\ref{eq:final_xi}), one can find that $\hat{t}_{c}$ is in the range of $2-2.5$. 
Thus a limit of $|\hat{t}|\leq 2$ is considered in this paper for the analytical solution to be used. Note that the downwind location where $|\hat{t}|=2$ is reached 
depends on inflow and turbine operating conditions as discussed 
in \S\ref{sec:results_laminar} and \S\ref{sec:turb_model}.

In the following, we develop an empirical relationship for $\hat{\xi}(\theta,\hat{t})$. Using the same harmonic terms as those in the analytical solution (\ref{eq:final_xi}),  we assume the empirical model is given by
\begin{dmath}\label{eq:xi_empirical}
 \hat{\xi}(\theta,\hat{t}) = 1
 -\alpha\left[c_1(\hat{t}) \cos 2 \theta + \left(c_2(\hat{t}) \upchi \sin 2 \theta +c_3(\hat{t}) \cos 3 \theta \right)+\left(c_4(\hat{t})\upchi^2 \cos 2 \theta+c_5(\hat{t})\upchi\sin 3 \theta +c_6(\hat{t})\cos 2 \theta+c_7(\hat{t}) \cos4 \theta\right)\right],
\end{dmath}
where $\alpha$ is a constant, and $c_1(\hat{t}),...,c_7(\hat{t})$ are time-dependant coefficients that need to be determined. Equation (\ref{eq:xi_empirical}) must provide predictions similar to the analytical solution (\ref{eq:final_xi}) at small values of $|\hat{t}|$, but it must provide desirable results for $|\hat{t}|$ beyond $\hat{t}_c$ too. To achieve this goal, we use a hyperbolic tangent function to express $c_i(\hat{t})$,
\begin{equation}
    c_i(\hat{t})=a_i\tanh{\frac{\hat{t}^{n_i}}{b_i}},
\end{equation}
where $i=1,...,7$ and $a_i$, $b_i$ and $n_i$ are constants. 
As $\hat{t}\to\infty$, (\ref{eq:xi_empirical}) is reduced to
\begin{dmath}\label{eq:xi_empirical_infty}
\lim_{\hat{t}\to\infty} \hat{\xi}(\theta,\hat{t}) = 1
 -\alpha\left[a_1\cos 2 \theta + \left(a_2\upchi \sin 2 \theta +a_3\cos 3 \theta \right)+\left(a_4\upchi^2 \cos 2 \theta+a_5\upchi\sin 3 \theta +a_6\cos 2 \theta+a_7\cos4 \theta\right)\right].
\end{dmath}
In (\ref{eq:xi_empirical_infty}), the overall curled shape of the wake is achieved by a suitable selection of $a_1,...,a_7$, while the extent of curling is controlled by $\alpha$. The constant $\alpha$ is introduced to ensure that the maximum possible curling is obtained as $\hat{t}\!\to\!\infty$. To estimate values of $a_1,..., a_7$, one can use the analytical solution at an arbitrary dimensionless time which is sufficiently large but still smaller than $\hat{t}_c$. By doing so, we construct the curled shape of the wake for $\hat{t}\to\infty$ from its analytical shape at a finite value of $\hat{t}$. The values of $a_i$ ($i=1,...,7$) are obtained based on the analytical solution (\ref{eq:final_xi}) at $\hat{t}=2$ and are written in Table \ref{table}, although other values of $\hat{t}$ could be used. 

In order to obtain maximum curling as $\hat{t}\!\to\!\infty$, $\hat{\xi}$ must go to zero at $\theta=\theta_0$, where $0\leq\theta_0\leq2\pi$, while $\hat{\xi}$ should be still non-negative for all polar angles. Note that for $\upchi=0$, $\theta_0$ is a multiple of $\pi$, but it might have a different value if $\upchi\neq 0$ due to the wake rotation. Based on the values of $a_i$ written in Table \ref{table}, one can show that $\hat{\xi}$ given in (\ref{eq:xi_empirical_infty}) is always equal or greater than $1-0.792\alpha\sec({0.33\upchi})$. Therefore, $\alpha$ must be equal to $1.263\cos(0.33\upchi)$ to guarantee that (i) $\hat{\xi}$ never becomes negative, and (ii) maximum possible curling occurs as $\hat{t}\to\infty$.
 \begin{table}
  \centering
  \begin{tabular*}{.6\columnwidth}{@{\extracolsep{\stretch{1}}}*{8}{c}@{}}\noalign{\vskip 1mm}
  Term& 1  & 2& 3    & 4 & 5 & 6 & 7\vspace{2 mm}\\\hline\noalign{\vskip 1mm}
        $a_i$ & $\frac{1}{2}$ & $-\frac{1}{3}$ & $-\frac{1}{4}$ &$-\frac{1}{6}$ & $\frac{5}{16}$  & -$\frac{5}{48}$ & $\frac{7}{48}$ \\\hline\noalign{\vskip 1mm}
      $b_i$ &  $4\alpha$& $8\alpha$ &  $8\alpha$ & $16\alpha$  & $16\alpha$ & $16\alpha$& $16\alpha$\\\hline\noalign{\vskip 1mm}
    $n_i$ & 2 & 3 &  3 &  4 & 4 & 4& 4\\\hline\noalign{\vskip 1mm}
  \end{tabular*}
  \caption{\small Coefficients of the empirical vortex sheet shape model (\ref{eq:xi_empirical}), where $\alpha=1.263\cos(0.33\upchi)$ and $c_i=a_i\tanh(\hat{t}^{n_i}/b_i)$.}\label{table}
\end{table}

\begin{figure}
 \centerline{\includegraphics[width=.95\textwidth]{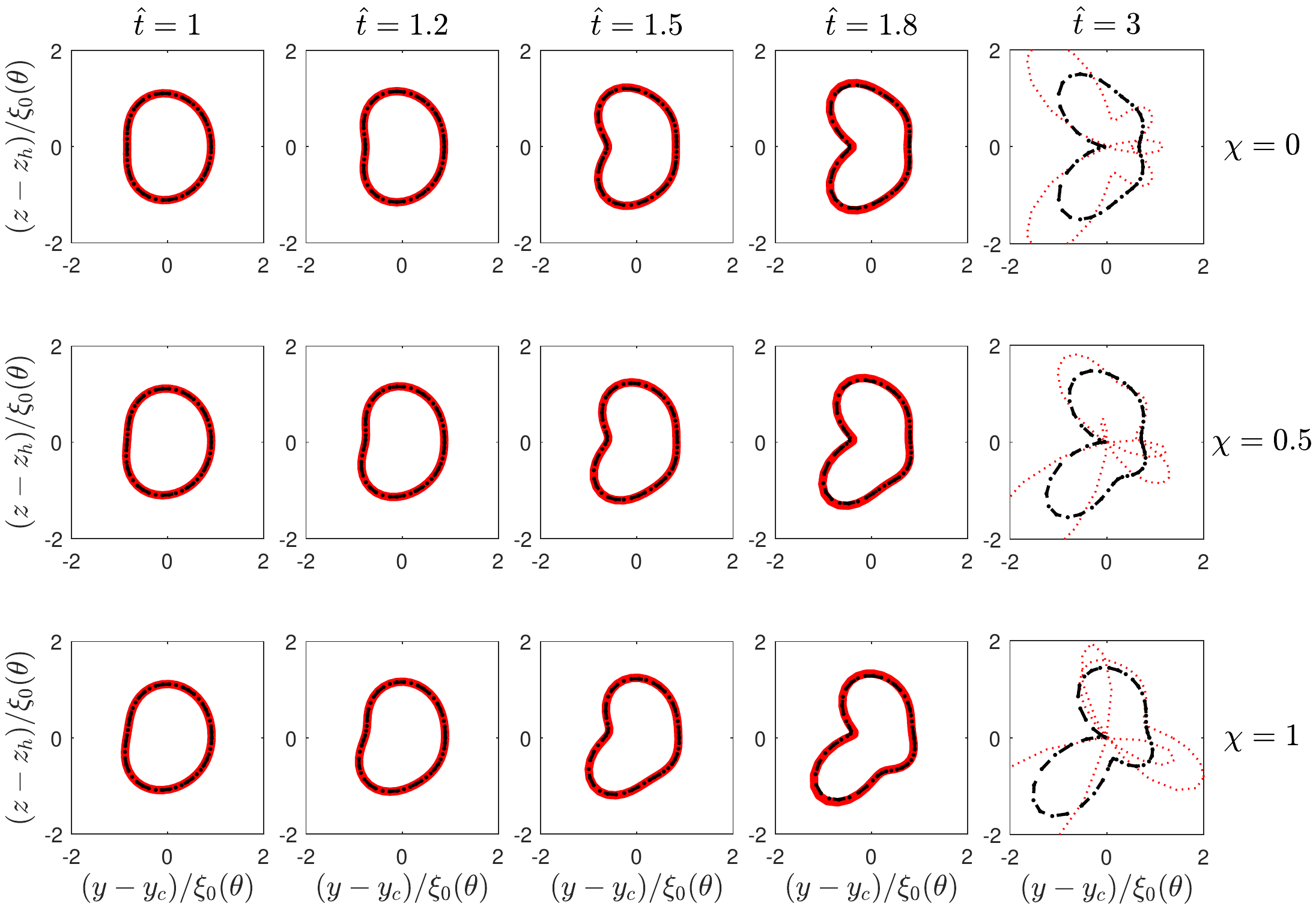}}
 \caption{The curled shape of the wake for different values of dimensionless time $\hat{t}$ and rotation rate $\upchi$. The analytical solution (\ref{eq:final_xi}) is shown by the red colour (solid curves for $\hat{t}\leq 2$ and dotted curves for $\hat{t}>2$), and the proposed empirical relation (\ref{eq:xi_empirical}) is shown by black dashed lines.}
 \label{fig:empirical_shape}
\end{figure}

The asymptotic behavior of (\ref{eq:xi_empirical}) at $\hat{t}\to\infty$ was used to find values of $a_i$ and $\alpha$. Next, we look into the asymptotic behaviour of (\ref{eq:xi_empirical}) at $\hat{t}\to0$ to find $b_i$ and $n_i$. The function $a_i\tanh{\hat{t}^{n_i}/b_i}$ asymptotes to $a_i\hat{t}^{n_i}/b_i$ as $\hat{t}\to0$. Therefore, we obtain:
    \begin{dmath}\label{eq:xi_empirical_zero}
 \lim_{\hat{t}\to0}\hat{\xi}(\theta,\hat{t}) = 1
 -\alpha\left[\frac{a_1\hat{t}^{n_1}}{b_1}\cos 2 \theta + \left(\frac{a_2\hat{t}^{n_2}}{b_2}\upchi \sin 2 \theta +\frac{a_3\hat{t}^{n_3}}{b_3}\cos 3 \theta \right)+\left(\frac{a_4\hat{t}^{n_4}}{b_4}\upchi^2 \cos 2 \theta+\frac{a_5\hat{t}^{n_5}}{b_5}\upchi\sin 3 \theta +\frac{a_6\hat{t}^{n_6}}{b_6}\cos 2 \theta+\frac{a_7\hat{t}^{n_7}}{b_7}\cos4 \theta\right)\right].
\end{dmath}
Equation (\ref{eq:xi_empirical_zero}) should match the analytical solution (\ref{eq:final_xi}). Therefore, values of $n_i$ and $b_i$ can be readily determined, and they are written in Table \ref{table}. Figure \ref{fig:empirical_shape} shows predictions of the wake shape based on both the analytical and empirical solutions for different values of dimensionless time $\hat{t}$ and rotation rate $\upchi$. As expected, the empirical relation provides similar results to those of the analytical solution at small values of $\hat{t}$. Moreover, unlike the analytical solution, it provides reasonable predictions for large values of $\hat{t}$ (i.e., $|\hat{t}|>2$). 

\section{Large Eddy Simulation code}
\label{AppC:LES}
The numerical code used in this study is the pseudo-spectral large-eddy simulation code LESGO. It solves the filtered Navier-Stokes equations in rotational form
\begin{align}
&\frac{\partial \tilde{u}_i}{\partial x_i} = 0 \\
&\frac{\partial \tilde{u}_i}{\partial t} +\tilde{u}_j \left( \frac{\partial \tilde{u_i}}{\partial x_j}  - \frac{\partial \tilde{u_j}}{\partial x_i} \right)=  - \frac{\partial \tau_{ij}}{\partial x_j}  - \frac{\partial \tilde{p}^*}{\partial x_i} -\frac{1}{\rho}\frac{\partial p_\infty}{\partial x} \delta_{i1} + f_i,
\end{align}
where $\tilde{u}_i$ is the filtered velocity field, $\tau_{ij}$ is the subgrid stress tensor, $\tilde{p}^*$ is the modified pressure, $\partial_x p_\infty$ is the driving streamwise pressure gradient, and $f_i$ are the turbine forces. The subgrid stress tensor is defined using an eddy viscosity model
\begin{equation}
\tau_{ij} = - 2 \nu_T \tilde{S}_{ij}
\end{equation}
where $\tilde{S}_{ij} = \frac{1}{2}\left( \partial_j \tilde{u}_i + \partial_i \tilde{u}_j\right)$ is the filtered strain rate tensor. LESGO simulates Cartesian domains using a psuedo-spectral numerical scheme that mixes spectral derivatives in the streamwise direction $x$ and spanwise direction $y$ with second-order finite-differencing in the vertical direction $z$. Time integration uses the second-order Adams-Bashforth method.

We consider wind turbines under both uniform laminar inflow~\citep{shapiro2018} and turbulent boundary layer inflows~\citep{shapiro2020decay}. In the uniform inflow simulations, the Smagorinsky model is used for the eddy viscosity
\begin{equation}
\nu_T = C_s^2 \Delta^2 \vert S \vert \qquad \qquad \vert S \vert^2 = 2 S_{ij} S_{ij}
\end{equation}
where $C_s=0.16$ is the Smagorinsky coefficient. This choice has a negligible effect on the results as the eddy diffusivity is confined to regions of the flow with strong velocity gradients. The inflow conditions are enforced via a fringe region at the end of the domain. This fringe region smoothly transitions to a prescribed inflow velocity $U_i(0,y,z) = U_{\rm{in}} \delta_{i1}$~\citep{Stevens2014a}. At the top and bottom of the domain, the stress-free boundary conditions for streamwise $u$ and spanwise $v$ velocity components are applied and the no-penetration condition is applied for the vertical component $w$.

For the turbulent boundary layer simulations, the subgrid stress tensor is modeled using the Lagrangian-averaged scale dependent model~\citep{Bou-Zeid2005a}. Turbulent inflow conditions are applied using a fringe region forcing where the inflow condition is sampled from a concurrently running simulation~\citep{Stevens2014a} with shifted periodic boundary conditions~\citep{Munters2016a}. At the ground, the stress with a roughness height of $z_0$ is modeled using the equilibrium wall model. The total wall stress is given by
\begin{equation}
\tau_w = - \left[  \frac{\kappa u_\tau}{\ln(\Delta z/ 2 z_0)} \right]^2,
\end{equation}
where $\Delta z$ is the vertical grid spacing, $\kappa$ is the von-K\'{a}rm\'{a}n constant, and $u_\tau$ is the velocity magnitude $u_\tau^2 = \tilde{u}(\Delta z/2)^2 + \tilde{v}(\Delta z/2)^2$.  The wall stress is then apportioned to each component
as
\begin{equation}
\left.\tau_{i3}\right\vert_{\text{wall}}= \tau_w \frac{\tilde{u}_i}{u_\tau}.
\end{equation}
LESGO implements the actuator disk model \black with rotation (ADM-R) \black using the local formulation for the thrust \black and angular \black forces. \black The thrust force in this forumulation is
\begin{equation}
    T = \frac{1}{2} \rho \pi R^2 C_T' U_d^2, 
\end{equation}
where $C_T'$ is the local thrust coefficient and $U_d$ is the disk averaged velocity. The total thrust force is distributed across the swept area of the disk using the filtered indicator function $\mathcal{R}(\boldsymbol{x})$
\begin{equation}
    \boldsymbol{f}(\boldsymbol{x}) = T \mathcal{R}(\boldsymbol{x}) \boldsymbol{n}
\end{equation}
in the unit normal direction to the disk $\boldsymbol{n} = \cos \beta \boldsymbol{i} + \sin \beta \boldsymbol{j}$. The filtered indicator function is found by convolving \begin{equation}
    \mathcal{R}(\boldsymbol{x}) = \int G(\boldsymbol{x}-\boldsymbol{x}') \mathcal{I}(\boldsymbol{x'}) \, d^3 x'
\end{equation} an indicator function $\mathcal{I}(\boldsymbol{x})$ for the geometric shape of the disk with finite thickness $s$ with a Gaussian filtering kernel $G(\boldsymbol{x})$ with a characteristic width $\sigma_R = 1.5 h / \sqrt{12}$ that is proportional to the mean grid size $h = \sqrt{\Delta x^2 + \Delta y^2 + \Delta z^2)}$.

To define the angular force \black of the rotation actuator disk \black in a local formulation, we consider the actuator disk model for unyawed wind turbines. The disk-averaged velocity $U_d = U_{\rm{in}} (1-a)$, streamwise velocity deficit $u_x = 2U_{\rm{in}} a$, and angular change in velocity $u_\theta = 2 a' \Omega r$ are defined based on the streamwise $a$ and tangential $a'$ induction factors and rotation rate of the disk $\Omega$. Considering annular rings of the swept area of the rotor, the annular thrust is the product of the annular flow rate and the change in momentum $\textrm{dT} = U_{\rm{in}} a(1-a) \pi r \textrm{d}r \rho 2 U_{\rm{in}}$. The annular torque is similarly the product of the annular flow rate and the change in angular momentum $\textrm{d}Q = U_{\rm{in}} (1-a) \pi r \textrm{d}r \rho 2 a' \Omega r^2$.

Since the annular power is equal to the products $\textrm{d}P = \textrm{d}T u_d $ and $\textrm{d}P = \textrm{d}Q \Omega$, the following equality
\begin{equation}
\frac{1}{2} \rho U_{\rm{in}}^3 4a(1-a)^2 \pi r \, \textrm{d}r = \frac{1}{2} \rho U_{\rm{in}}^3 4 a'(1-a) \left(\frac{\Omega r}{U_{\rm{in}}} \right)^2 \pi r \, \textrm{d}r
\end{equation}
can be solved for the tangential induction factor, so we obtain
\begin{equation}
    a' =  a(1-a)\left(\frac{U_{\rm{in}}}{\Omega r}\right)^2.
\end{equation}

The torque can be written in local and standard forms 
\begin{align}
\textrm{d}Q = \frac{1}{2} \rho U_{\rm{in}}^2 R 4a(1-a)^2 \frac{U_{\rm{in}}}{\Omega R} \pi r \textrm{d}r = \frac{1}{2} \rho U_{\rm{in}}^2 R C_Q \pi r \textrm{d}r = \frac{1}{2} \rho u_d^2 R C_Q' \pi r \textrm{d}r.
\end{align}
Defining the tip speed ratio $\lambda = \Omega R/U_{\rm{in}}$ and local tip speed ratio as $\lambda' = \Omega R/u_d$ the torque coefficient is 
\begin{equation}
    C_Q' = \frac{C_P'}{\lambda'}.
\end{equation}
For ideal actuator disks $C_P' = C_T'$. The tangential force is then found from $r \textrm{d} f_\theta = \textrm{d} Q$
\begin{equation}
    \textrm{d} f_\theta = \frac{1}{2}\rho U_{\rm{in}}^2 \frac{C_P}{\lambda}\frac{R}{r} \pi r dr  = \frac{1}{2}\rho u_d^2 \frac{C_P'}{\lambda'}\frac{R}{r} \pi r \textrm{d}r.
\end{equation}
The tangential force applied by the rotating rotor in LESGO is written as 
\begin{equation}
    \boldsymbol{f}_\theta(\boldsymbol{x}) = \frac{1}{2} \rho \pi R^2 \frac{C_T'}{\lambda'} u_d^2 \frac{R}{r} \mathcal{R}(\boldsymbol{x})\hat{\boldsymbol{\theta}},
\end{equation}
where $\hat{\boldsymbol{\theta}}$ is the tangential unit vector. \black In the case of no wake rotation, the tip speed ratio is infinite and the angular force vanishes. \black

\bibliographystyle{jfm}
\bibliography{main, reference_Majid}

\end{document}